\documentclass[12pt]{article}

\usepackage{txfonts}
\usepackage[]{graphicx}
\usepackage{booktabs}
\usepackage{subcaption}
\usepackage{xcolor}
\usepackage[normalem]{ulem} 

 \def\mso{\,\mathrm{M}_\odot}

 \def\kms{\, \mathrm{km}\, {\mathrm s}^{-1}}

  \def\simle{\mathrel{\hbox{\rlap{\hbox{\lower4pt\hbox{$\sim$}}}\hbox{$<$}}}}
 \def\simgr{\mathrel{\hbox{\rlap{\hbox{\lower4pt\hbox{$\sim$}}}\hbox{$>$}}}}

\newcommand*\xbar[1]{%
   \hbox{%
     \vbox{%
       \hrule height 0.5pt 
       \kern0.5ex
       \hbox{%
         \kern-0.1em
         \ensuremath{#1}%
         \kern-0.1em
       }%
     }%
   }%
}

\usepackage{ulem}
\usepackage{color}

\newenvironment{sciabstract}{%
\begin{quote} \bf}
{\end{quote}}

\title{Stellar mergers as the origin of the blue main-sequence band in young star clusters}
\author{Chen Wang$^{1,10}$, Norbert Langer$^{1,2}$, Abel Schootemeijer$^{1}$, Antonino Milone$^{3}$,\\ Ben Hastings$^{1,2}$, Xiao-Tian Xu$^{1,2}$, Julia Bodensteiner$^{4,5}$, Hugues Sana$^{4}$, Norberto Castro$^{6}$, \\
D. J. Lennon$^{7,8}$, Pablo Marchant$^{4}$, A. de Koter$^{9,4}$, Selma E. de Mink$^{10,9,11 }$\\
\\
\footnotesize{$^{1}$Argelander-Institut f\"ur Astronomie, Universit\"at Bonn, Auf dem H\"ugel 71, 53121 Bonn, Germany}\\
\footnotesize{$^{2}$ Max-Planck-Institut f\"ur Radioastronomie, Auf dem H\"ugel 69, 53121 Bonn, Germany} \\
\footnotesize{$^{3}$Dipartimento di Fisica e Astronomia ``Galileo Galilei'', Univ. di Padova, Vicolo dell'Osservatorio 3, I-35122 Padova, Italy}\\
\footnotesize{$^{4}$Institute of Astronomy, KU Leuven, Celestijnlaan 200D, 3001 Leuven, Belgium}\\
\footnotesize{$^{5}$European Southern Observatory, Karl-Schwarzschild-Str. 2 85738 Garching bei München, Germany}\\
\footnotesize{$^{6}$Leibniz-Institut f\"ur Astrophysik Potsdam (AIP), An der Sternwarte 16, 14482 Potsdam, Germany}\\
\footnotesize{$^{7}$ Instituto de Astrof\'isica de Canarias, E-38200 La Laguna, Tenerife, Spain}\\
\footnotesize{$^{8}$ Dpto. Astrof\'isica, Universidad de La Laguna, E-38205 La Laguna, Tenerife, Spain}\\
\footnotesize{$^{9}$Astronomical Institute ``Anton Pannekoek", University of Amsterdam, Science Park 904, 1098 XH Amsterdam, The Netherlands}\\
\footnotesize{$^{10}$Max Planck Institute for Astrophysics, Karl-Schwarzschild-Strasse 1, 85748 Garching, Germany}\\
\footnotesize{$^{11}$Center for Astrophysics, Harvard-Smithsonian, 60 Garden Street, Cambridge, MA 02138, USA}\\
\\
\footnotesize{$^\ast$To whom correspondence should be addressed; E-mail:  cwang@astro.uni-bonn.de. cwang@mpa-garching.mpg.de}}
\date{\today}

\begin{document} 


\baselineskip24pt


\maketitle 
\bigskip


\begin{sciabstract}
Recent high-quality Hubble Space Telescope (HST) photometry shows that the main sequences (MS) stars of young star clusters form two discrete components in the color-magnitude diagram (CMD). Based on their distribution in the CMD, we show that stars of the blue MS component can be understood as slow rotators originating from stellar mergers. We derive the masses of the blue MS stars, and find that they follow a nearly flat mass function, which supports their unusual formation path. 
Our results imply that the cluster stars gain their mass in two different ways, by disk accretion leading to rapid rotation, contributing to the red MS, or by binary merger leading to slow rotation and populating the blue MS.
We also derive the approximate merger time of the individual stars of the blue MS component, and find a strong early peak in the merger rate, with a lower level merger activity prevailing for tens of Myr. This supports recent binary formation models, and explains new velocity dispersion measurements for members of young star clusters. Our findings shed new light on the origin of the bi-modal mass, spin, and magnetic field distributions of main-sequence stars.
\end{sciabstract}


The main sequence (MS) of star clusters is a cornerstone of stellar formation and evolution \cite{1967ARA&A...5..571I,2018A&A...616A..10G}. 
In the last decade, the simple picture of star clusters as an ensemble of coeval stars born
with identical initial conditions has been challenged. Old and very massive globular clusters host multiple stellar populations, with differences in chemical compositions\cite{2012A&ARv..20...50G,2017MNRAS.464.3636M}.  
Recent Hubble Space Telescope (HST) observations have also revealed that the MSs of young open star clusters (with ages between $\sim$ 15\,Myr and $\sim$ 600\,Myr) are composed of several discrete components\cite{2009A&A...497..755M,2016MNRAS.458.4368M,2017ApJ...844..119L,2018MNRAS.477.2640M,2019ApJ...876...65L}, but with identical chemical composition \cite{2016MNRAS.458.4368M,2014ApJ...793L...6M}.
In the color-magnitude diagram (CMD), this is characterised by a split MS from the vicinity of the turn-off all the way to a faint magnitude, with the red MS component containing more stars than the blue MS (e.g. NGC 1755 in Fig.\,1).
The emergence of a sub-population of blue MS stars is ubiquitous in Magellanic Cloud clusters younger than $\sim$600\,Myr \cite{2018MNRAS.477.2640M}.

In this paper, we use the $\sim 60$\,Myr old Large Magellanic Cloud (LMC) cluster NGC\,1755  as an example to investigate the origin of its split MS, with three more clusters, that are NGC\,1818, NGC\,2164 in the LMC and NGC\,330 in the Small Magellanic Cloud (SMC), discussed in Supplementary Information A. 
The MSs in these clusters all show a distinct blue component in the CMD. In NGC\,1755, the color difference between the blue component and the well-defined red (major) MS is up to 0.25 magnitudes \cite{2018MNRAS.477.2640M} (see Fig.\,1a). 
The lower portion of the blue MS is the narrowest, while the upper portion is more diffuse.
The blue MS stars comprise roughly 20\% of the cluster stars.

\begin{figure}[ht]
\centering
\includegraphics[width=\linewidth]{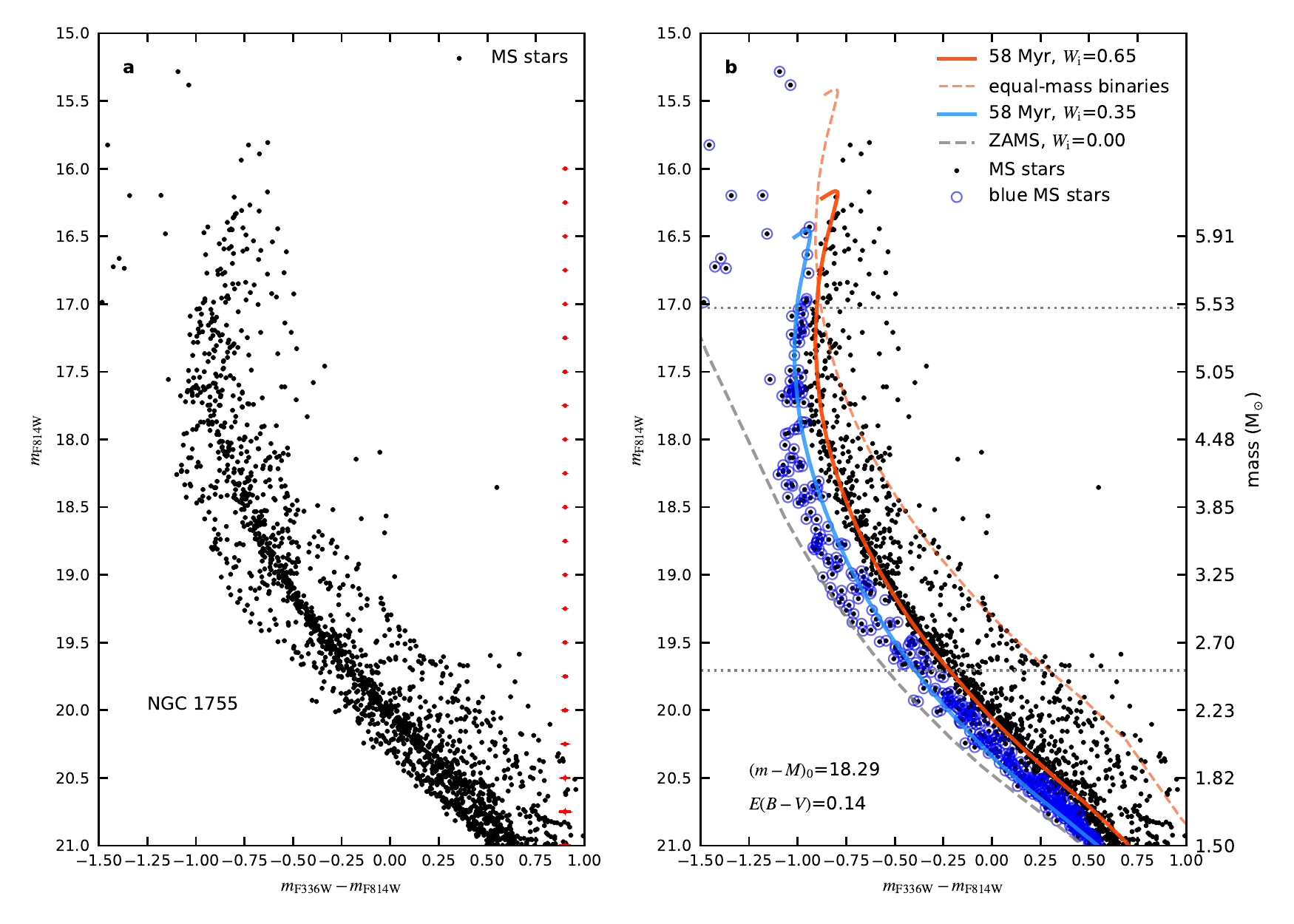}
\caption{Color-magnitude diagram of NGC\,1755 and the isochrone fits. {\bf a}: Color-magnitude diagram of the stars (black dots) in the main-sequence region of the Large Magellanic Cloud open star cluster 
NGC\,1755 based on high-quality HST photometry \cite{2018MNRAS.477.2640M}. Typical 1$\sigma$ errors at different magnitudes are shown with red error bars on the right. {\bf b}: Isochrone fit for the red (major) main sequence
of NGC\,1755, signified by the densest stellar concentration,   
using stellar models with a rotation parameter of $W_{\rm i}=0.65$ (red solid line), and identification
of the blue main-sequence stars (blue circles; cf., Supplementary Information A). 
The solid blue line represents the corresponding isochrone of single star models with $W_{\rm i}=0.35$.
Here, $W_{\rm i}=v_{\rm rot,i}/v_{\rm crit,i}$ is the ratio between rotational velocity and break-up velocity at the zero-age main sequence. 
The red dashed line shows the positions of the equal-mass binaries in which both components have $W_{\rm i}=0.65$. The grey dashed line indicates the zero-age main-sequence line of non-rotating stars. The adopted age, distance modulus $(m-M)_0$ and reddening $E(B-V)$ are indicated. 
The stars between the two thin horizontal dotted lines are used in our mass function analysis (see Supplementary Information D). The right y-axis 
shows the stellar mass according to the mass-magnitude of the fast rotating stellar models.
}
\end{figure}

There is well-founded observational evidence that supports rotation being responsible for the split MS, with blue MS stars rotating significantly slower than other cluster stars (designated here as red MS stars) \cite{Bastian+2009,2017ApJ...846L...1D,2018AJ....156..116M,2018MNRAS.477.2640M,2018MNRAS.480.3739B,Sun_2019,2020MNRAS.492.2177K}. In particular, the spectroscopically measured average projected rotation velocity of red MS stars in NGC\,1818 was found to be $202\pm 23\kms$, while it was only $71\pm 10 \kms$ for the blue MS stars in this cluster \cite{2018AJ....156..116M}.
While previous studies suggest that extremely rapid rotation ($\sim90\%$ of initial critical rotation) of the red MS stars may be required to account for the color difference between blue and red MS \cite{2017NatAs...1E.186D,2018MNRAS.477.2640M},
our stellar models show that adopting $\sim$65\% of initial critical rotation for the red MS stars
and $\sim$35\% for the blue MS stars provides indeed a good fit to the observed red MS and 
to the best discernible part of the blue MS (Fig.\,1b and Supplementary Information B, see Supplementary Information A on how we identify blue MS stars). 
This also agrees with the currently available, although sparse
spectroscopic rotational velocity measurements \cite{2018AJ....156..116M,Sun_2019,2020MNRAS.492.2177K}, and, notably, with the bi-modal distribution of the rotation rates of B and A-type field MS stars, in which the average projected rotational velocities of the slow and fast components are $\sim 20-100\kms$ and $180-250\kms$, respectively \cite{Huang+2010,2012A&A...537A.120Z,2013A&A...550A.109D}. 

Other scenarios for the origin of blue MS stars have been proposed.
Suggestions that the blue MS stars formed in a second burst of star formation after the formation of the majority of the cluster stars
\cite{2011ApJ...737....4G,2017MNRAS.467.3628C} are in conflict with the persisting
color difference between blue and red MS even far below the turn-off, since the faintest stars are essentially unevolved.
Similarly, the
trend that the apparent cluster age spread measured from the MS widths near the turn-off increases for older Magellanic Cloud and Galactic star clusters cannot be explained by a second star burst, 
and is related to effects of rotation
\cite{2015A&A...575A..62N,2015MNRAS.453.2070N,2018ApJ...869..139C}.

It has further been proposed that the blue MS stars were born with similar rotational velocities as the red MS stars,
but that their rotation has slowed down subsequently due to tidal braking \cite{2017NatAs...1E.186D}. 
Whereas tidal braking of close binaries provides a viable spin-down mechanism, it does not produce enough blue MS stars, since most close binaries will appear redder than non-rotating single stars regardless of the rotation rate of their components, due to the presence of two stars in the observed point source. Only very low-mass ratio binaries containing slow rotators are expected to contribute to the blue MS stars.
The blue MS has also been suggested to originate from a combination of MS stars with He-star companions and stellar mergers, where the MS split is explained
by a bi-modal distribution of post-merger masses \cite{2018ApJ...860..132Y}.
Bi-modal disk-locking during the star formation process has also been suggested to explain the 
observed MS dichotomy\cite{2020MNRAS.495.1978B}.
In this model, the blue MS stars are slow rotators due to a longer disks-locking time during their pre-MS accretion phase, compared to the red MS stars. While this may reproduce the rotation dichotomy, it cannot explain that closer to the turn-off, more and more of the blue MS stars are located to the blue side of the blue MS isochrone.

Binary mergers offer a natural way to create the blue MS population. A binary merger creates a star that is more massive than either of its progenitor stars, with a core hydrogen content that is higher than that of an equally old single star
of the same mass. Thus, merger products may have the same
age as all other cluster stars, but appear younger in the CMD, signified by their bluer color. 
Previous studies have shown that tight binary stars which merged as a consequence of the expansion 
of their component stars during hydrogen burning evolution form the blue stragglers that are brighter and bluer than the turn-off stars in star clusters \cite{2015ApJ...805...20S,2020ApJ...888L..12W}.
The continuity in the CMD between the blue stragglers and the fainter blue MS stars displayed in several clusters (NGC\,1866, NGC\,1856, NGC\, 294, KRON\,34, with ages between 200\,Myr to 500\,Myr \cite{2018MNRAS.477.2640M}) provides us a 
further clue for the merger origin of the latter.

Further evidence for a merger origin of the blue MS stars is provided by analysing their mass functions. 
We fit the CMD positions of blue and red MS stars, i.e., the magnitude of the individual stars
along the constructed isochrones, with single star models of the appropriate spin and age to obtain 
a measurement of the stellar mass (Supplementary Information D). 
We then fit the mass distributions of both groups of stars with power laws.
For the red MS stars, the derived power law exponent of $\gamma\simeq -2.17\pm 0.15$ is close to that of a Salpeter law ($\gamma=-2.35$) in the mass range of $5.5\dots 2.5\mso$. In the same range, the mass function of the blue MS stars, is found to follow a power law with an exponent of $\gamma= -1.03\pm 0.32$, representing a much shallower mass distribution (Fig.\,2). 
Similar results are obtained for the other clusters investigated (Supplementary Information D).
This indicates that the blue MS stars more massive than $2.5\mso$ are not formed by the same mechanism as the red MS stars,
just at a later time, but that both groups emerge from different formation mechanisms.
The shallow slope of the mass function of the blue MS stars is in fact consistent with their merger origin, as it may be the result of the observed decreasing close binary fraction with decreasing stellar mass \cite{2015ApJ...805...20S}. This 
is also expected according to recent binary formation models \cite{2020MNRAS.491.5158T}. 

\begin{figure}[ht]
\centering
\includegraphics[width=\linewidth]{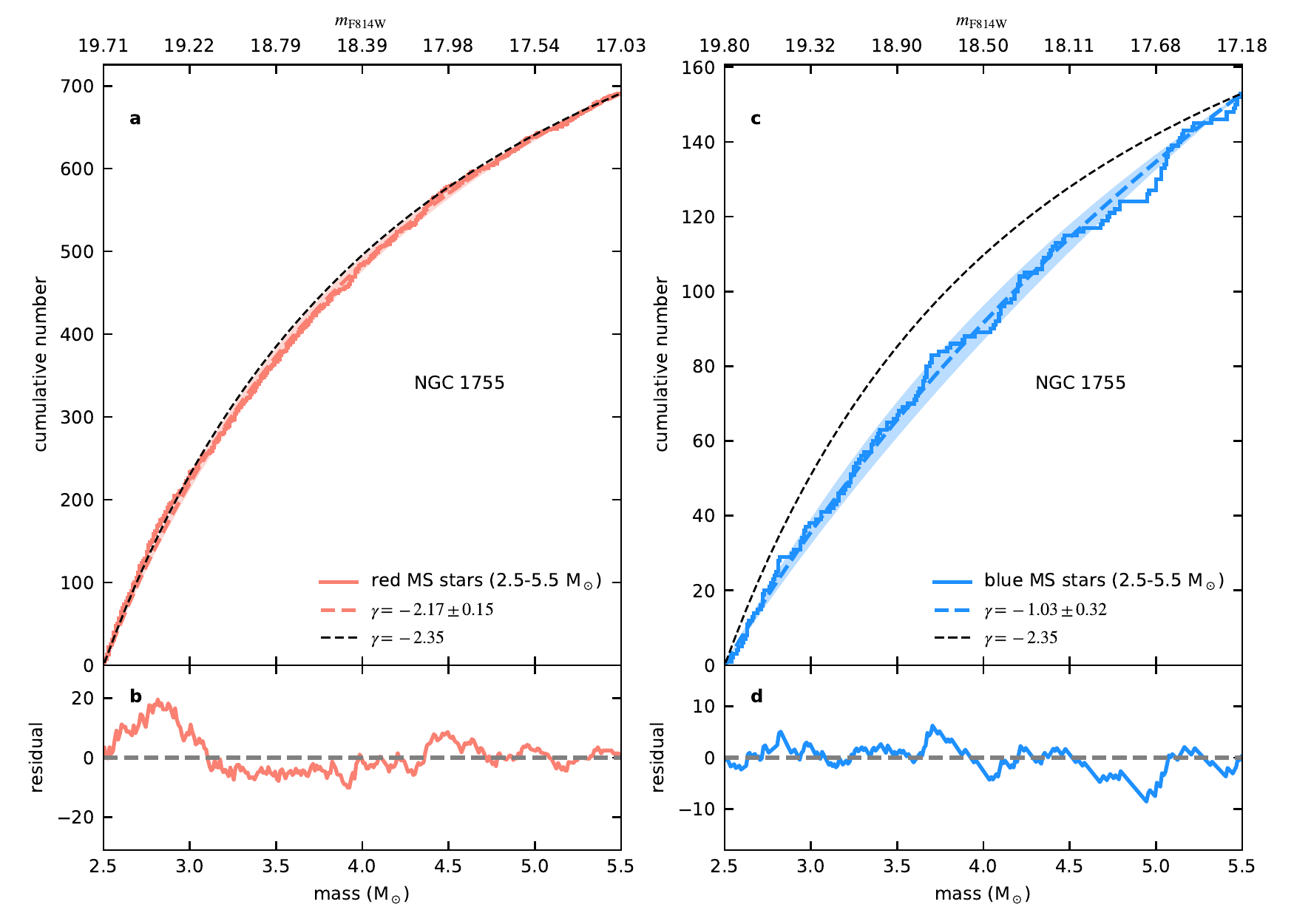}
\caption{Mass function of the red and the blue main-sequence stars in NGC\,1755. {\bf a} and {\bf c}: Cumulative number distribution of the red main-sequence (red lines) and the blue main-sequence stars (blue lines) more massive than $2.5\mso$ in NGC\,1755
as a function of their mass, as derived from the stellar model isochrones.
Dashed lines display distributions according to the best-fitting power-law mass function.
The shading reflects one sigma errors of our power-law fitting, and correspond to the error of the power law indices 
given in the legend. The dashed black lines show the distributions predicted by a power law with index $\gamma = -2.35$ 
(the traditional Salpeter IMF), assuming the same total number of stars as the corresponding number in the observed populations. 
We show the apparent magnitudes for given masses on the top of the figures.
{\bf b} and {\bf d}: Residuals, i.e., difference between the colored solid and dashed lines in Panels\,a and\,c, as a function of mass,
with a residual of zero indicated by the grey dashed lines.
}
\end{figure}

A merger origin of the blue MS stars may also hold the clue for their slow rotation \cite{2018AJ....156..116M,2020MNRAS.492.2177K}.
While initially, a stellar binary merger contains a large angular momentum surplus due to the
orbital angular momentum, recent hydrodynamic binary merger calculations show that
the merger product loses most of that in a puffed-up stage right after the merger,
such that it settles as a slow rotator on the MS after a Kelvin-Helmholtz time scale \cite{2019Natur.574..211S}. 
These simulations suggest that large-scale magnetic fields form in the merger product, such that a further
spin-down due to magnetic wind braking may occur subsequently.

Unlike previously discussed scenarios, the merger origin of the blue MS stars provides a coherent explanation of the dichotomies in color,
rotation, and mass function slope,
and of the peculiarly wider
blueward extension of the blue MS
for brighter magnitudes. It leads us to the exciting conclusion that
stars come to accumulate their mass in two fundamentally different ways. On the one hand, the majority of stars form by accretion of gas via accretion disks, which is the dominant path, leading to a well populated red MS with a rotation rate of slightly larger than half of critical rotation. On the other hand, a fair fraction of the so created stars merge with a similar mass companion 
and produce a blue MS star, rotating significantly slower and obeying a different initial mass function than the red MS stars. This bi-modality in the star formation process may therefore be at the root of the observed bi-modalities in stellar spins \cite{2012A&A...537A.120Z,2013A&A...550A.109D}, magnetic fields \cite{2009ARA&A..47..333D}, and mass functions \cite{2018MNRAS.477.2640M}, and of course location in the CMD. 
In the following, we shall discuss these aspects in more detail.

The narrowness of the red MS (Fig.\,1a) implies that potentially, accretion induced star formation may result in a rather 
narrow distribution of MS rotational velocities in the considered mass range, with a peak near 65\% of critical rotation, which is in agreement with considerations of gravitational torques between stars and disks \cite{2011MNRAS.416..580L}, and also in agreement with the spectroscopic velocity measurements of both the red MS stars in young star clusters \cite{2018AJ....156..116M,Sun_2019,2020MNRAS.492.2177K} and the B and A-type field stars \cite{Huang+2010,2012A&A...537A.120Z,2013A&A...550A.109D}. 
Furthermore, there is evidence for stars in binaries being born with very similar spins as single stars \cite{2015A&A...580A..92R}.
Whereas the information about the initial spins of the stars which merged to become blue MS stars is wiped out in the merger process,
these stars appear to rotate so slowly that their color is largely unaffected by rotation (Supplementary Figure\,6). 
While we see a significant number of MS stars with extreme rotation, most notably the Be stars, which are likely 
evolved MS stars which are spun-up either by mass transfer from a binary companion \cite{2020ApJ...888L..12W} 
or as single stars by their contracting cores \cite{2008A&A...478..467E,2020A&A...633A.165H}, it is thus conceivable 
that upper MS stars are formed either rotating slowly, or about half critically.

In Fig.\,3, we show the distribution of our detailed binary evolution models (see Supplementary Information C) at 30\,Myr in the CMD. The two star components have SMC-like metallicity and are assumed to rotate at 55\% of their critical velocities at their zero-age MSs. 
Without further assumptions, our detailed binary models reproduce all the MS components observed in the SMC star cluster NGC\,330. We find that binaries consisting of a MS star and a hot stripped companion star, and short period binaries with tidally spun-down components only account for an insignificant fraction of the observed blue MS stars (see Supplementary Information C and Supplementary Figure\,8). 

\begin{figure}[ht]
\centering
\includegraphics[width=\linewidth]{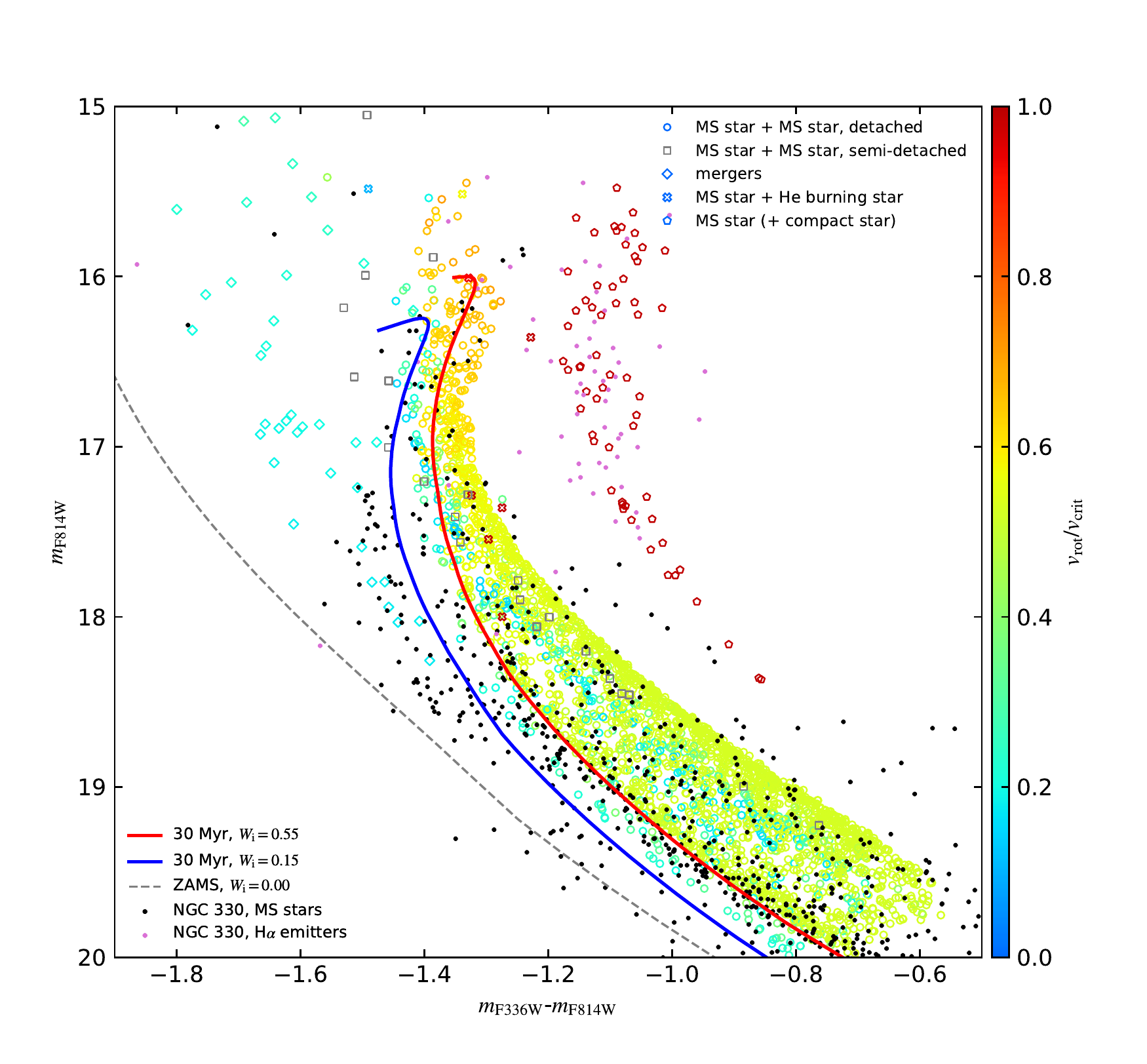}
\caption{Binary evolution models at 30\,Myr in the color-magnitude diagram. Each open symbol indicates a binary model (or a binary merger product), showing the combined magnitude and color of the two stars (or the magnitude and color of the binary merger product). Circles indicate detached binary models containing two main-sequence stars, while squares correspond to semi-detached binary models containing two main-sequence stars. Diamonds designate main-sequence merger products. Crosses correspond to binary models containing a main-sequence star and a stripped helium burning star, while pentagons denote main-sequence star models whose companion has evolved to a compact object and may have left them as a consequence of supernova kick. The semi-detached systems are marked in grey, while the color for other open symbols shows the current rotational velocity of either the visually brighter component in a binary model or of a binary merger product. The observed main-sequence stars and H$\alpha$ emitters are overplotted with small black and purple dots, respectively. Distance modulus and reddening are assumed as Supplementary Figure\,7b.
The isochrones and zero-age main-sequence line are the same as those in Supplementary Figure\,7b.
}
\end{figure}
     
However, Fig.\,3 shows that binary evolution driven by the nuclear timescale expansion of the individual stellar components
cannot account for the large number of observed blue MS stars, particularly far below the turn-off, since stellar expansion starts very slowly during hydrogen burning. 
Therefore, a merger fraction of 
the order of 20\% all along the MS can only be produced by decaying binary orbits. 
In fact, there are multiple lines of evidence for this.
Tidal forces imposed by the circum-binary matter from which the stars formed 
are known to induce a drastic decay of the binary orbit\cite{2012A&A...543A.126K}.
Recent binary formation models indeed predict about 30\% of binary B\,stars to merge
during the pre-MS evolution or shortly thereafter \cite{2020MNRAS.491.5158T}. Orbit decay is also required to
explain the high observed fraction of very close massive binary systems \cite{2012Sci...337..444S}.
Direct evidence for pre-MS binary orbit decay is 
provided from observations of pre-MS binaries in a nearby star forming
region\cite{2017A&A...599L...9S}. 

To quantify the consequences of stellar mergers, we construct simple merger models following the scheme of \cite{2016MNRAS.457.2355S}. The strongest apparent 
rejuvenation of merger products, as measured by time difference between the cluster age and the apparent age of the merger product identified through single star models, occurs in binaries with a mass ratio of one (Supplementary Information E). Fitting models of equal-mass mergers to an individual 
blue MS star in the CMD therefore allows us to obtain a lower limit to its merger time, i.e., the moment in the cluster history when it formed through the coalescence of its progenitor binary. Supplementary Figure\,14 shows lines of constant merger time for equal mass binaries, indicating the current positions of merger products which were created at the
indicated times, in the CMD of NGC\,1755. These lines extend from the blue MS,
for a merger time, $t_\mathrm{merge}$, corresponding to the time of cluster formation (defined here as $t=0$), almost all the way to the zero age
MS for a merger time close to the cluster age ($t=58\,$Myr). Notably, the lines of constant merger time cover most blue MS stars.
Blue MS stars to the red side of the line for $t_\mathrm{merge}=0$ may have a slightly faster 
initial rotation than $W_\mathrm{i}=0.35$ (see Supplementary Figure\,6). We note that this applies in 
particular to blue MS stars below $\sim 19.5\,$mags, or $\sim 2.5\,$M$_{\odot}$, which is analogous to
the disappearance of the slowest rotators in Galactic field stars \cite{2012A&A...537A.120Z}.

Assuming equal-mass binary mergers, we can constrain the history of stellar merger events in NGC\,1755 and other clusters. 
The unknown exact rotation rate of the blue MS stars remains the dominant error source in deriving their merger time from their positions in the CMD. We therefore derive the merger time of each blue MS star 
for different assumptions on the rotational rate (see Fig.\,4b and Supplementary Information E), 
which yields representative uncertainties of the merger times.
For stars fainter than $\sim 19\,$mags, we cannot constrain the merger times any more. 
Figure\,4a shows the result of integrating over Gaussian merger time probability distributions 
for each star, with the shaded area indicating the $1\sigma$ bootstrapping envelope (see Supplementary Information E).
Figure\,4 implies that the merger rate was largest within the first few Myrs of the cluster evolution, after which
it dropped considerably. However, on a reduced but substantial level, the merging activity prevailed 
for several tens of Myr.
Since by adopting equal-mass binary mergers, we only obtain lower limits to merger time of each star, 
the true merger times might be somewhat larger than implied by Fig.\,4. 
However, varying the mass ratio of the pre-merger binaries only has a limited effect (see Supplementary Information E). 
In fact, it is the relative distance between each blue MS star to the blue MS isochrone that determines its merger time. That the star density is the highest in the vicinity of this isochrone unambiguously implies an early peak in the merger rate.
In addition, we find a moderate positive correlation between merger time and stellar magnitude (see Supplementary Information E), which is consistent with recent binary formation simulations that suggest binaries with higher masses to merge earlier than binaries with lower masses \cite{2012A&A...543A.126K}.
We repeated the analysis for the three other clusters (see Supplementary Information E)
and found very similar results.

\begin{figure}[ht]
\centering
\includegraphics[width=0.90\linewidth]{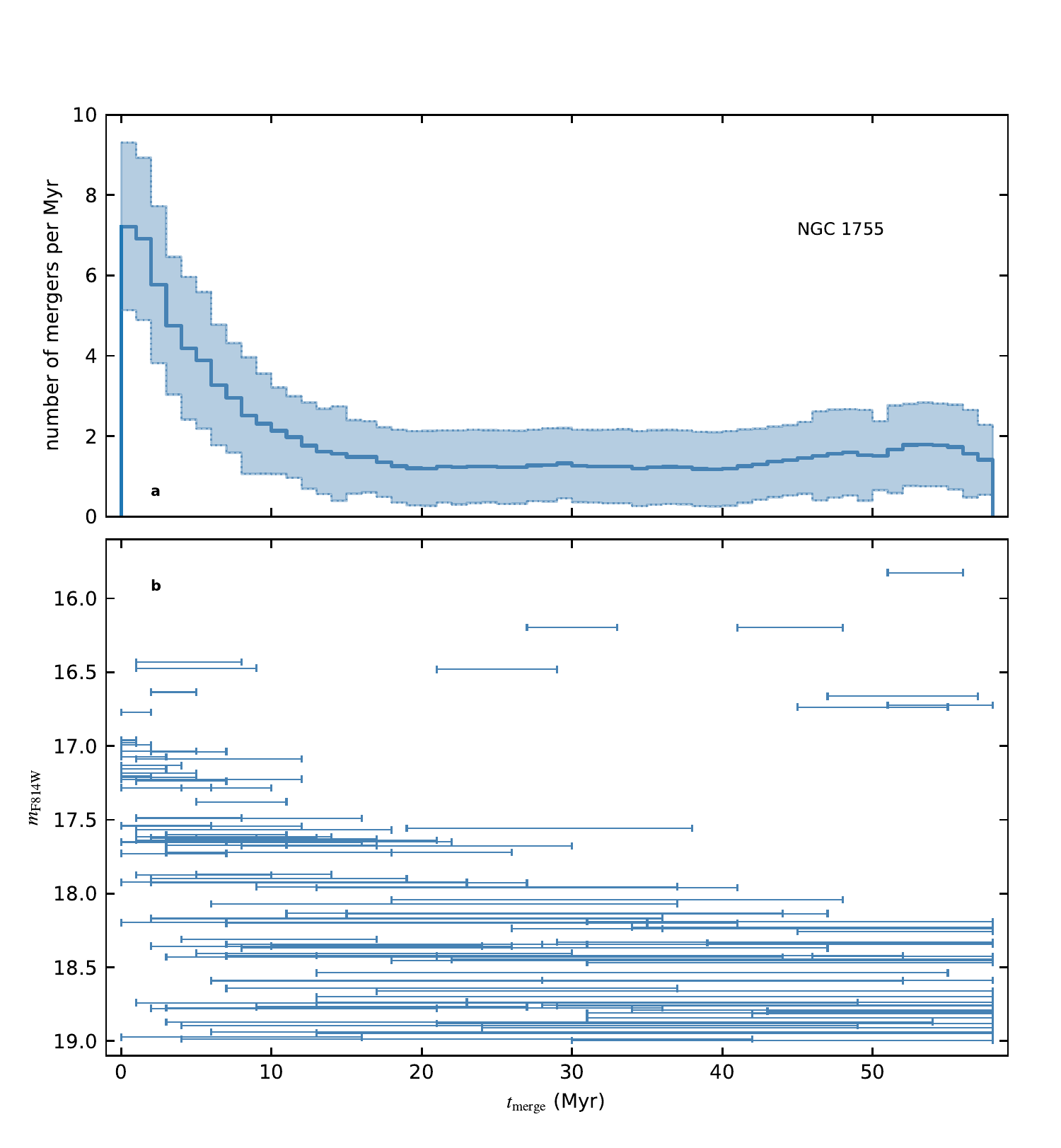}
\caption{Stellar merger history in NGC\,1755. {\bf a} (upper panel): Merger rate, expressed as number of merger events per Myr, 
as function of time, as derived for NGC\,1755.
This histogram is obtained by taking in to account a Gaussian probability distribution 
of the merger rotation over the time range displayed in Panel\,b for each star, and
summing up their contributions in each time bin of width 1\,Myr. The shaded area indicates bootstrapped $1\sigma$ estimates.
{\bf b} (bottom panel): Magnitude versus merger time of the blue main-sequence stars in NGC\,1755
(blue horizontal error bars). Magnitude errors are small and are not shown here (see Fig.\,1). 
The upper and lower limits of the merger time are derived from the minimum and maximum
rotation velocities consistent with the color, individually for each star.
A time of 58\,Myr corresponds to today.
}
\end{figure}

The strong peaks at early time in our derived merger histories appear consistent with the observed rapid rise of the velocity dispersions in star clusters during their
first few million years of evolution \cite{2021A&A...645L..10R}. Notably, this timescale corresponds
to the duration of the pre-main sequence phase of stars in the considered mass range, during which time they are
bloated and thus more prone to tidal effects.

The derived merger activity on a timescale  at least ten times longer is more difficult to understand. 
The similarity of this timescale to the timescale of violent relaxation
of star clusters, on which they re-virialise after expelling the gas left over after star formation ended,
suggests that dynamical processes may play a role. While stellar encounters can lead to binary hardening,
the stellar density in the investigated clusters is too small to render this process efficient \cite{2011MNRAS.413.1810B}.
It appears more likely that triple and higher-order multiple system can foster the merging of
their inner binary components, e.g. via Lidov-Kozai cycles \cite{1962AJ.....67..591K,2001ApJ...562.1012E} or by passing stars \cite{2018AJ....156...96W} that can cause a high eccentricity of the inner binaries and trigger the subsequent tidal friction. 
Indeed, we show in Supplementary Information E and Supplementary Figure\,20,
that even today, after most of the mergers may have occurred, the CMDs of the discussed 
young star clusters bear evidence for a current fraction of triple and higher order multiple systems of at least of several percent,
which may imply that the corresponding merger activity is still ongoing.

Binary mergers of MS stars have also been suggested to be responsible for the generation of the large scale B-field found in about 10\% of the upper MS stars \cite{2009MNRAS.400L..71F},
and magnetohydrodynamic (MHD)-simulations of the merger process appear to support this idea \cite{2019Natur.574..211S}. According to our analysis,
the fraction of merger stars ($\sim 20$\%) is larger than the observed fraction
of magnetic stars \cite{2013MNRAS.429..398P}. This means that either not every merger event leads to a magnetic star, or that the merger generated B-fields decay on a timescale comparable to the nuclear time scale of the stars. The fact that the topological requirement of intertwined toroidal and poloidal field components 
\cite{2004Natur.431..819B} is not a guaranteed outcome of the turbulent merger phase provides evidence for the former scenario, while the distribution of the fractional MS ages of magnetic massive stars 
\cite{2016A&A...592A..84F} justifies the latter.

Furthermore, our results have implications for the understanding of the stellar initial mass function (IMF).
They imply that the IMF when measured from field stars consists of two components with largely different slopes.
In star clusters, on the other hand, the stellar mass function is not static as often assumed, 
but it may evolve as the merger rate changes with time, offering a challenging but feasible way to test the proposed interpretation.

In our analysis, we consider clusters younger than 100\,Myr.  However, our conclusion that orbit decay and binary mergers account for the slow rotators may apply to much older clusters. Effects of rotation are found in clusters of up to 2\,Gyr \cite{2009A&A...497..755M,2018MNRAS.477.2640M,2020MNRAS.492.2177K}, below which the information on the stellar birth spin and on large scale B-fields is erased by the star's convective envelopes, which produce their own magnetic activity and lead to magnetic spin-down.

\section*{Method}
\subsection*{Single star models}
We use the detailed one-dimensional stellar evolution code MESA\cite{Paxton2011,Paxton2013,Paxton2015,Paxton2019}. Most of the physical assumptions are identical to those utilized in \cite{2011A&A...530A.115B}. 
The exception is that we use a mass-dependent overshooting parameter $\alpha_\mathrm{ov}$ (i.e., the number of pressure scale-heights by which the hydrogen-burning core is extended).
For an initial mass of 20\,M$_\odot$ we use $\alpha_\mathrm{ov} = 0.3$ \cite{2011A&A...530A.115B}. Below that, $\alpha_\mathrm{ov}$ decreases linearly such that it reaches a value of $\alpha_\mathrm{ov} = 0.1$ at 1.66\,M$_\odot$ \cite{2016A&A...592A..15C}. Below 1.66\,M$_\odot$, $\alpha_\mathrm{ov}$ has an even steeper linear decrease such that it equals zero at 1.3\,M$_\odot$ (where the convective core disappears). This mass-dependence accounts for the trend that the width of the distribution of field MS stars in the CMD increases with mass \cite{2014A&A...570L..13C,2016A&A...592A..15C,2019A&A...625A.132S}. 
The adopted $\alpha_\mathrm{ov}$ in this work are similar to the findings in \cite{2021A&A...648A.126M}.
We emphasize here that although overshooting still remains poorly constrained, and it can affect the location of the turn-off stars in the CMD, the uncertainty of overshooting does not play a role in explaining the observed double MS \cite{2017ApJ...836..102Y}.
We include differential rotation, rotationally induced internal mixing, magnetic angular momentum transport, stellar wind mass loss, and non-equilibrium CNO nucleosynthesis. 
We use the standard mixing-length theory to model convective mixing with a mixing-length parameter $\alpha_{\mathrm{MLT}}=1.5$. The Ledoux criterion is used to determine the boundaries of convective zones. 
In the superadiabatic layers that are stable according to the Ledoux criterion but unstable according to the Schwarzschild criterion, we assume that semiconvection occurs with a mixing parameter of $\alpha_\mathrm{SC}=10$ \cite{2019A&A...625A.132S}. We model rotational mixing as a diffusive process\cite{2000ApJ...528..368H}, taking into account the effects of dynamical and secular shear instabilities, the Goldreich-Schubert-Fricke instability \cite{1967ApJ...150..571G,1968ZA.....68..317F}, and the Eddington-Sweet circulations \cite{1925Obs....48...73E}.  The efficiency parameter of rotational mixing is $f_\mathrm{c}=1/30$ as proposed in \cite{1992A&A...253..173C}. We include the Tayler-Spruit dynamo for the transport of angular momentum\cite{Spruit2002,Heger2005}.

We follow the mass-loss recipe used in \cite{2011A&A...530A.115B}: for hydrogen-rich stars with  surface hydrogen mass fraction $X_{\mathrm{s}} \geq 0.7$, the wind prescription of \cite{2001A&A...369..574V} is used, while for hydrogen-poor stars with $X_{\mathrm{s}} \leq 0.4$, the WR mass-loss prescription of \cite{Hamann1995} is used. For intermediate surface hydrogen abundances $0.4 < X_{\mathrm{s}} < 0.7$, we linearly interpolate the value of $\mathrm{log} \,\dot{M}$ between the two prescriptions. The metallicity-dependent stellar winds scale as $\dot{M} \propto Z^{0.85}$ \cite{2001A&A...369..574V}.

We consider both LMC and SMC metallicity, with $Z_{\rm LMC}=0.00484$ and $Z_{\rm SMC}=0.00218$ \cite{2011A&A...530A.115B}. We compute single star models in a mass range of 1.05$\mso$ and 20 $\mso$ in a dense grid with $\Delta \log m_{\rm i} = 0.02$, from the zero-age MS, which is defined at the position where 3\% of hydrogen is burnt to avoid the initial model relaxation, until well beyond core hydrogen exhaustion.
We define a stellar model's initial fractional critical rotation as $W_{\rm i} = v_{\rm rot,i}/v_{\rm crit,i}$, where $v_{\rm rot, i}$ and $v_{\rm crit, i}$ are its average surface rotational velocity and its break-up velocity at the zero-age MS , respectively. We follow MESA definition of the critical velocity $v_{\rm crit} = \sqrt{Gm/R_{\rm eq}}$, where $m$ and $R_{\rm eq}$ are the mass and equatorial radius of the stellar model, respectively.
Then for each mass, we compute model sequences for $W_{\rm i}$ ranging from 0.15 to 0.75, in intervals of 0.1,
as well as non-rotating models. We did not go higher than that because we encounter numerical problems when computing stellar models with $W_{\rm i}>0.75$.

To construct the stellar distribution in the CMD, we first calculate the absolute magnitude of a star in a given filter F as $M_\mathrm{F}=M_{\mathrm{bol}}-BC_{\rm F}$, where $M_{\mathrm{bol}}=M_{\mathrm{bol,\,\odot}}-2.5\,{\mathrm{log\,}}(L/L_{\odot})$ is the bolometric magnitude of a star, $BC_{\rm F}$ its bolometric correction for the adopted filter, and $M_{\mathrm{bol,\,\odot}}=4.74$ mag\cite{2016AJ....152...41P}. The bolometric correction is obtained by interpolating tables computed from 1D atmosphere models based on ATLAS12/SYNTHE \cite{1993sssp.book.....K,1970SAOSR.309.....K}
for either the HST/WFC3 F814W or F336W filters as these correspond to observations used in this paper. The absorption coefficients are $A_{F814W}=2.04\,E(B-V)$, $A_{F336W}=5.16\,E(B-V)$ \cite{2018MNRAS.477.2640M}, where $E(B-V)$ is the reddening. The apparent magnitude is then obtained by $m_\mathrm{F}=M_\mathrm{F}+A_\mathrm{F}+(m-M)_0$, where $(m-M)_0$ is the distance modulus. 

\subsection*{Binary star models}
These newly computed binary models are an extension of the binary models in \cite{2020ApJ...888L..12W}, using MESA version 8845, but assuming that both binary components start with 55\% of their critical rotation velocities at the zero-age MS. The physics assumptions adopted in each star model otherwise are the same as in the single star models of this work. We use SMC-like metallicity for our binary models, because the metallicity-dependent stellar wind is weak, such that any differences between binary and single star models are mainly caused by binary interaction. 
We briefly describe the physics and assumptions adopted for the binary interactions in the following. 

We simultaneously compute the detailed structure of both components, together with the orbital evolution. We assume the orbit to be circular. Our binary models are not synchronized initially but have 55\% of their critical velocities ($W_{\mathrm i}=0.55$), as we have shown in the last section that single stars with this velocity are consistent with the observed red MS of young star clusters. 
The two stars in each binary model can exchange mass and angular momentum via Roche lobe overflow. The mass transfer rate is implicitly adjusted such that the radius of the donor star is restricted to its Roche lobe radius \cite{1999A&A...350..148W}. The specific angular momentum accreted by the secondary star depends on whether the accretion is ballistic or occurs via a Keplerian disk. If the orbit of a binary system is wide enough to avoid tidal spin-down, the accretor can reach critical rotation by accreting only a few percent of its initial mass. When this happens, we enhance the mass-loss rate of the accretor \cite{1998A&A...329..551L,Paxton2015} such that it remains rotating just below critical. We assume radiation as the driving force of the enhanced wind.
If the required mass loss is beyond the radiative capability of the system, we assume that the binary is engulfed in the excess material, and merges as a consequence. The merger models are computed with the method described in detail in the next section. We assume the merger products to have an initial spin of 15\% of critical rotation ($W_{\mathrm i}=0.15$), which is consistent with the blue MS when  single star models with $W_{\mathrm i}=0.55$ are used to reproduce the red MS.

We use a Monte Carlo scheme to generate the initial parameters of 3500 binaries, representing a cluster of $7.7\times 10^4 \mso$, with a binary fraction of 1 and the least star mass being 0.8$\mso$.
The initial primary mass varies from 3$\mso$ to 100$\mso$, following a Salpeter IMF with an exponent of -2.35, while the initial mass ratio ranges from 0.1 to 1, obeying a flat distribution.
The initial period varies from a minimum value, at which the two stars would encounter Roche lobe overflow at zero-age MS to 3162 days, following a flat distribution in logarithmic space. 

We follow the evolution of these binaries from the zero-age MS to core carbon exhaustion. If the core mass of the primary star exceeds the Chandrasekhar mass at the time of carbon depletion, we assume that a supernova explosion happens, and compute the remaining evolution of its companion as single star. Such a system produces either a binary system containing a MS star and a compact object or a single MS star, depending on whether the system remains bound after supernova kick, which is not calculated in our work.

\subsection*{Stellar merger models}
We follow the method in \cite{2016MNRAS.457.2355S} to compute models of the merger product of two MS stars. 
There, it is assumed that the chemical structure of the merger star adjusts itself to that of an ordinary single star with the
appropriate mass and age. Whereas the details of the internal mixing process
during a stellar merger event are clearly more complex, the major aspect of our simplified models
is confirmed by multi-dimensional merger simulations and detailed follow up calculations \cite{2013MNRAS.434.3497G,2019Natur.574..211S,2020MNRAS.495.2796S}.
These studies show that the convective core mass of the merger product increases to a mass 
found in single star models of the post-merger stellar mass. This increase in convective core mass brings fresh hydrogen to the center of the star which is responsible for the rejuvenation process.

We assume that all stars are born with moderate rotation (i.e., $W_{\rm i}=0.65$, see Supplementary Information B for the reason), and use 
corresponding single star models to obtain the masses $M_1$ and $M_2$ of the two stars in a binary system immediately before the merger. 
The mass of the merger product $M$ is expressed as 
$$M = (1-\Phi )\,(M_1+M_2),$$ 
where $\Phi =0.3\,q/(1+q)^2$ with $q=M_2/M_1$ describes the fraction of the mass lost by the binary in the merger event \cite{2013MNRAS.434.3497G} . 
We assume that the lost material has the same composition as the initial composition of the two stars. We then compute the average 
hydrogen mass fraction of the two stars $\xbar{X_1}$ and $\xbar{X_2}$ immediately before the merger. The hydrogen mass after and before the merger are connected through
$$M\, \xbar{X} =M_1\, \xbar{X_1} +M_2\, \xbar{X_2} -(M_1+M_2)\, \Phi \, X_0 ,$$
where $\xbar{X}$ is the average hydrogen mass fraction of the merger product and $X_0$ is the initial hydrogen mass fraction of the stars. 
We use our single star models to identify the one which has the same mass and average hydrogen mass fraction as the merger product, and treat it as the starting model of the merger product evolution. 
The age of this starting model denotes the apparent age $t_{\rm app}$ of the merger product immediately after the coalescence.
We then follow the evolution of the merger product until the required age (i.e., the age of the cluster). This means that we evolve the merger model further for a time equal to the difference between the cluster age and the age at which the coalescence happens.

The rotation rate of the merger products, which has not been constrained well to date, plays the dominant role in affecting their positions in the CMD at a given merger time.
Even though results of MHD simulations have suggested mergers to be slow rotators \cite{2019Natur.574..211S}, seldom faster-rotating stars are detected among blue stragglers \cite{2020MNRAS.492.2177K}. Current available velocity measurements of the blue stragglers report $v\sin i$ values to be between 20 and 270 $\kms$, significantly smaller than those of most other cluster members \cite{2005MNRAS.358..716S,2021AJ....161...37R}.
In Supplementary Figure\,6, we see that when the red MS is fitted with $W_{\rm i} = 0.65$, the vast majority of the blue MS stars correspond to slow rotators with $W_{\rm i} \simle 0.55$. Based on the rotational velocities considered in our single star model grids, we build grids of merger models with birth fractional critical rotational rates of 0, and from 0.15 to 0.55, in intervals of 0.1.

\section*{Data Availability}
The observational data in this work can be found at https://doi.org/10.5281/zenodo.5770868. The MESA inlist files used to compute the single and binary star models in this work can be downloaded at: https://doi.org/10.5281/zenodo.5233209.

\section*{Acknowledgments}
The research leading to these results has received funding from the European Research Council (ERC) under the European Union's Horizon 2020 research and innovation programme (grant agreement numbers 772225: MULTIPLES). 
CW acknowledges funding from CSC scholarship. 
This work has received funding from the European Research Council (ERC) under the European Union's Horizon 2020 research innovation programme (Grant Agreement ERC-StG 2016, No 716082 ``GALFOR'', PI: Milone, http://progetti.dfa.unipd.it/GALFOR). APM acknowledges support from MIUR through the FARE project R164RM93XW SEMPLICE (PI: Milone) and the PRIN program 2017Z2HSMF (PI: Bedin).
HS and JB acknowledge support from the FWO Odysseus program under project G0F8H6N.
NC gratefully acknowledges funding from the Deutsche Forschungsgemeinschaft (DFG) - CA 2551/1-1. 
DJL acknowledges support from the Spanish Government
Ministerio de Ciencia, Innovaci\'on y Universidades through grants PGC-2018-091 3741-B-C22 and from the Canarian Agency for Research, Innovation and Information Society (ACIISI), of the Canary Islands Government, and the European Regional Development Fund (ERDF), under grant with reference ProID2017010115.
PM acknowledges support from the FWO junior postdoctoral fellowship No.
12ZY520N. SdM acknowledges funding by the Netherlands Organization for Scientific Research (NWO) as part of the Vidi research program BinWaves with project number 639.042.728.

\section*{Author contributions}
CW, AS, BH, X-TX performed the stellar evolution calculations, based on earlier work by PM and his advice. NL worked out the analysis interpretation of the results, together with CW, AS, BH and SdM. AM, JB, HS, NC, DL, AdK provided an interpretation of the related observations. All authors reviewed the manuscript.

\section*{Correspondence}
Correspondence and requests for materials should be addressed to Chen Wang (email:cwang@astro.uni-bonn.de).
\section*{Competing Interests}
The authors declare no competing interests.



\section*{Supplementary Information}
\section*{A: Procedure to identify blue main-sequence stars}\label{suppsec:A}
The main characteristic feature in the distribution of the cluster stars in the CMD, which is the basis for our work,
is that the MS band is split into two distinct components. This is a striking feature not only visible in the CMD of MS stars in NGC\,1755, but rather a common feature detected in Magellanic Cloud star clusters younger than $\sim$ 600\,Myr \cite{2018MNRAS.477.2640M}. In this work, we fix our attention on clusters younger than 100\,Myr, whose distinct components can be compared with our models of stars exceeding 2\,$\mso$. Currently, high-quality data exists for six clusters younger than 100\,Myr in the LMC and the SMC (NGC\,330, NGC\,1818, NGC\,1805, NGC\,1755, NGC\,1850 and NGC\,2164), five of which are shown to exhibit the split MS (excluding NGC\,1805), but one of them (NGC\,1850) is composed of two sub-clusters \cite{2018MNRAS.477.2640M}. We investigate all the remaining four clusters in this work, which are the LMC clusters NGC\,1755, NGC\,1818, NGC\,2164 and the SMC cluster NGC\,330.
We use the observational data published in \cite{2018MNRAS.477.2640M} which includes corrections for differential reddening. We use the same method as in that work to eliminate the contamination from the foreground and background field stars.

In this section, we illustrate the procedure to distinguish the blue and the red MS stars in young star clusters. We use a model-independent method similar to \cite{2016MNRAS.458.4368M}, that star classification is based on their split distribution in the CMD. We calculate the color difference between each observed star and a fiducial line that best describes the well-populated red MS in the CMD. To determine this fiducial line, we first draw a line along the red MS by visual inspection. Then we smooth this line by selecting a sample of red MS stars whose color distances to this line are smaller than four times the photometric error at corresponding magnitudes. We divide the selected red MS stars into bins of 0.2\,mag. The final fiducial line is then determined by the median color and magnitude of the selected red MS stars in each magnitude bin (see the red line in Supplementary Figure\,1a for NGC\,1755). We draw the fiducial line from a magnitude below the cluster turn-off magnitude because the red and the blue MS stars are not well discernible near the turn-off.

Taking NGC\,1755 as an example, in our considered area (Supplementary Figure\,1b), we calculate the color difference $\Delta(m_\mathrm{F336W}-m_\mathrm{F814W})$ between each star and the fiducial line (Supplementary Figure\,1c). Supplementary Figure\,1d shows the histogram of $\Delta(m_\mathrm{F336W}-m_\mathrm{F814W})$ distribution in eight magnitude bins, with a bin size of 0.5\,mag. We only display the results for stars with $\Delta(m_\mathrm{F336W}-m_\mathrm{F814W})\leq 0.5$. Finally, we perform a bi-Gaussian fitting for $\Delta(m_\mathrm{F336W}-m_\mathrm{F814W})$ distribution. We remind the reader that this bi-Gaussian fitting depends somehow on which histograms are taken into account. To exclude the effect from potential binaries, in our bi-Gaussian fitting, we only include the histograms with $\Delta (m_\mathrm{F336W}-m_\mathrm{F814W})$ smaller than a critical value, which is 0.12 if $m_\mathrm{F814W} \geq 20$, 0.1 if $19.5 \leq m_\mathrm{F814W} < 20$, 0.08 if $19 \leq m_\mathrm{F814W} < 19.5$, and 0.06 if $17 \leq m_\mathrm{F814W} < 19$, taking into account the fact that $\Delta(m_\mathrm{F336W}-m_\mathrm{F814W})$ of a binary system with a specific mass ratio increases with magnitude. The dark grey and light grey histograms in Supplementary Figure\,1d represent the histograms included and excluded in the bi-Gaussian fitting, respectively. As to the normalization, we first normalize the dark grey histograms to an area of unity, which facilitates the bi-Gaussian fitting. Then we normalize the light grey histograms such that the ratio between the area of the light grey histograms and the dark grey histograms equals the number ratio of the stars excluded and included in the bi-Gaussian fitting.

We classify all stars with $\Delta(m_\mathrm{F336W}-m_\mathrm{F814W})$ smaller than the value at which the two Gaussian curves cross as blue MS stars. 
While at $m_\mathrm{F814W} \leq 17$, we use eye inspection to identify blue MS stars according to isochrone fitting (Supplementary Information B). The final identified blue MS stars in NGC\,1755 are indicated by the blue circles in Fig.\,1b. The same procedure is used to identify the blue MS stars in NGC\,330 (Supplementary Figure\,2), NGC\,1818 (Supplementary Figure\,3) and NGC\,2164 (Supplementary Figure\,4). In NGC\,330, we adopt a smaller $\Delta(m_\mathrm{F336W}-m_\mathrm{F814W})$ boundary in our bi-Gaussian fitting for the bright stars, which is 0.04 if $17.5 \leq m_\mathrm{F814W} < 18$ and 0.02 if $17 \leq m_\mathrm{F814W} < 17.5$.

\newcommand{\beginsupplement}{
\setcounter{figure}{0}
\renewcommand{\figurename}{Supplementary Figure}}
\beginsupplement

\begin{figure}[ht]
\centering
\includegraphics[width=\linewidth]{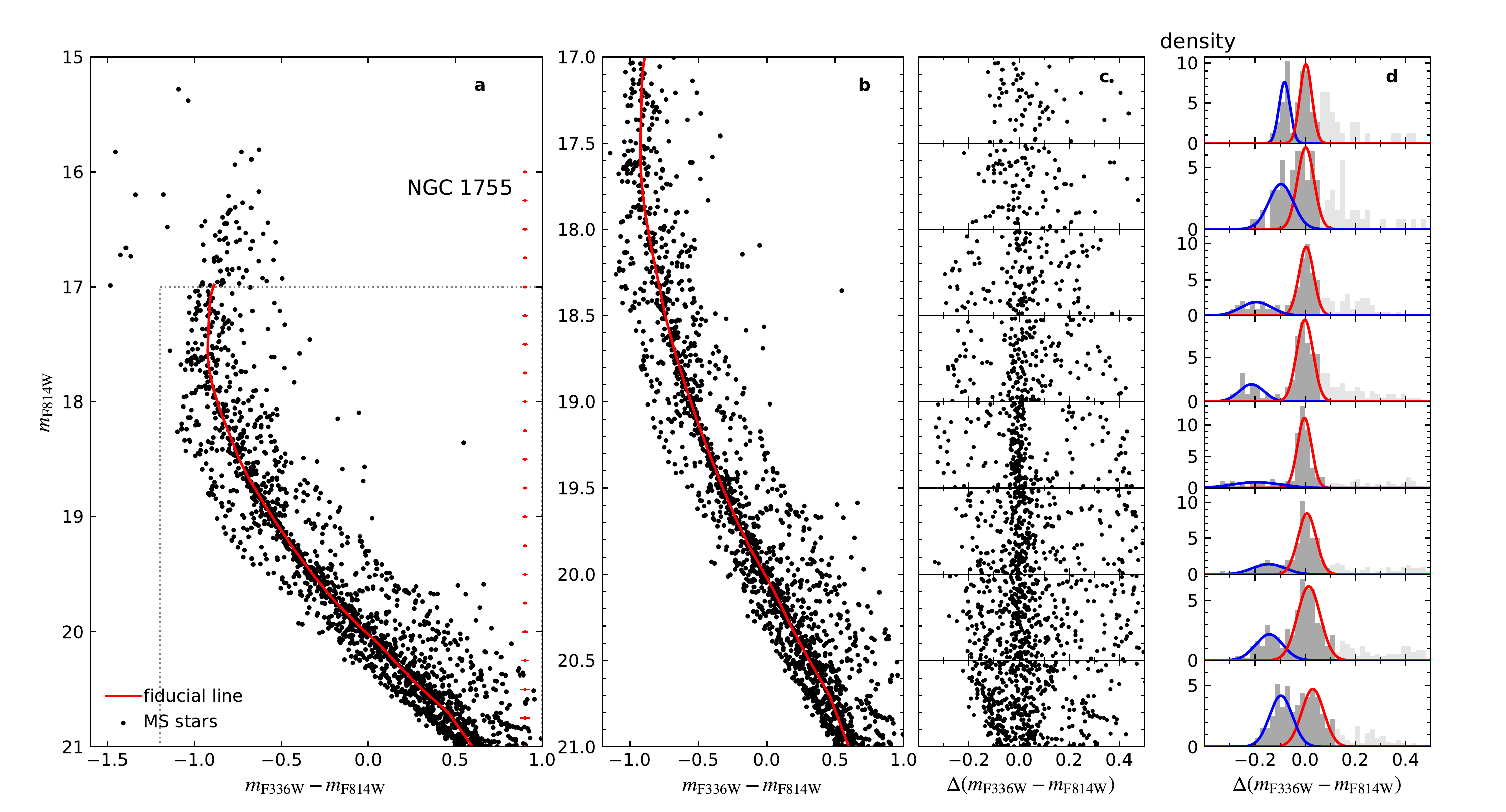}
\caption{Distinguishing red and blue main-sequence stars in NGC\,1755. 
{\bf a}: Same as Fig.\,1a, but with a fiducial line (solid red line) that best describes the distribution of the red main-sequence stars. The stars in the region delineated by the grey dotted lines are considered in this procedure. The red error bars on the right indicate 1$\sigma$ error at corresponding magnitudes.
{\bf b}: Zoom-in image of the area delineated by the grey dotted lines in panel a. {\bf c}: Horizontal color distance $\Delta(m_\mathrm{F336W}-m_\mathrm{F814W})$ of each star in panel b to the fiducial line as a function of magnitude. {\bf d}: Histogram distribution of the points in panel c in corresponding magnitude intervals. The dark grey (light grey) histograms are included (excluded) in our bi-Gaussian fitting. The dark grey histograms are normalized such that their total area equals one, while the light grey histograms are normalized such that their total area equals the number ratio of the stars excluded and included in our bi-Gaussian fitting.
The red and blue curves depict the components of our best bi-Gaussian fitting for the dark grey histograms. The stars with $\Delta(m_\mathrm{F336W}-m_\mathrm{F814W})$ smaller than that of the cross of the red and blue curves are identified as blue MS stars.
}
\label{Suppfig:n1755_split}
\end{figure}    

\begin{figure}[ht]
\centering
\includegraphics[width=\linewidth]{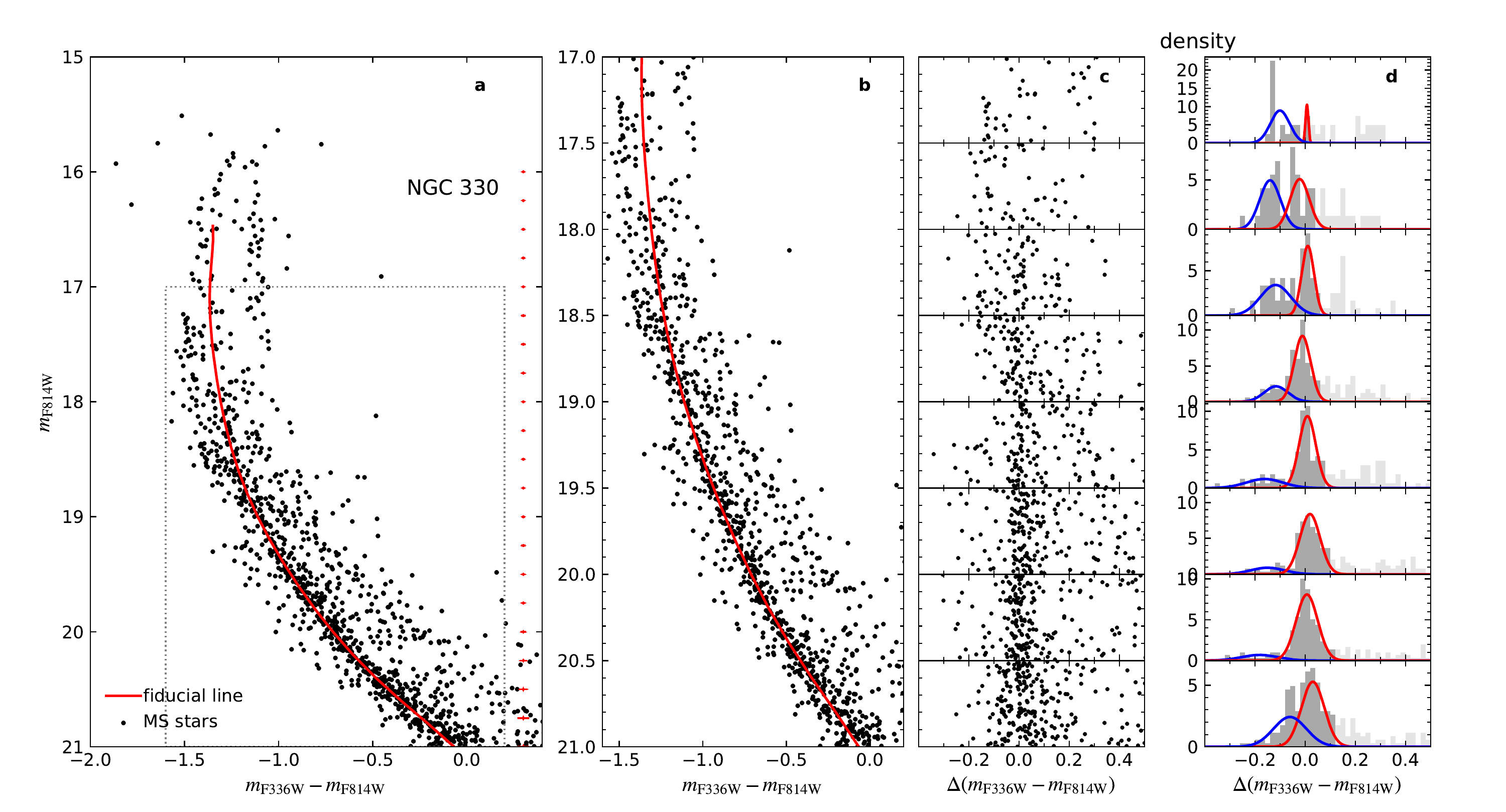}
\caption{Same as Supplementary Figure\,1, but for distinguishing the red and the blue main-sequence stars in NGC\,330.}
\label{Suppfig:n330_split}
\end{figure}    

\begin{figure}[ht]
\centering
\includegraphics[width=\linewidth]{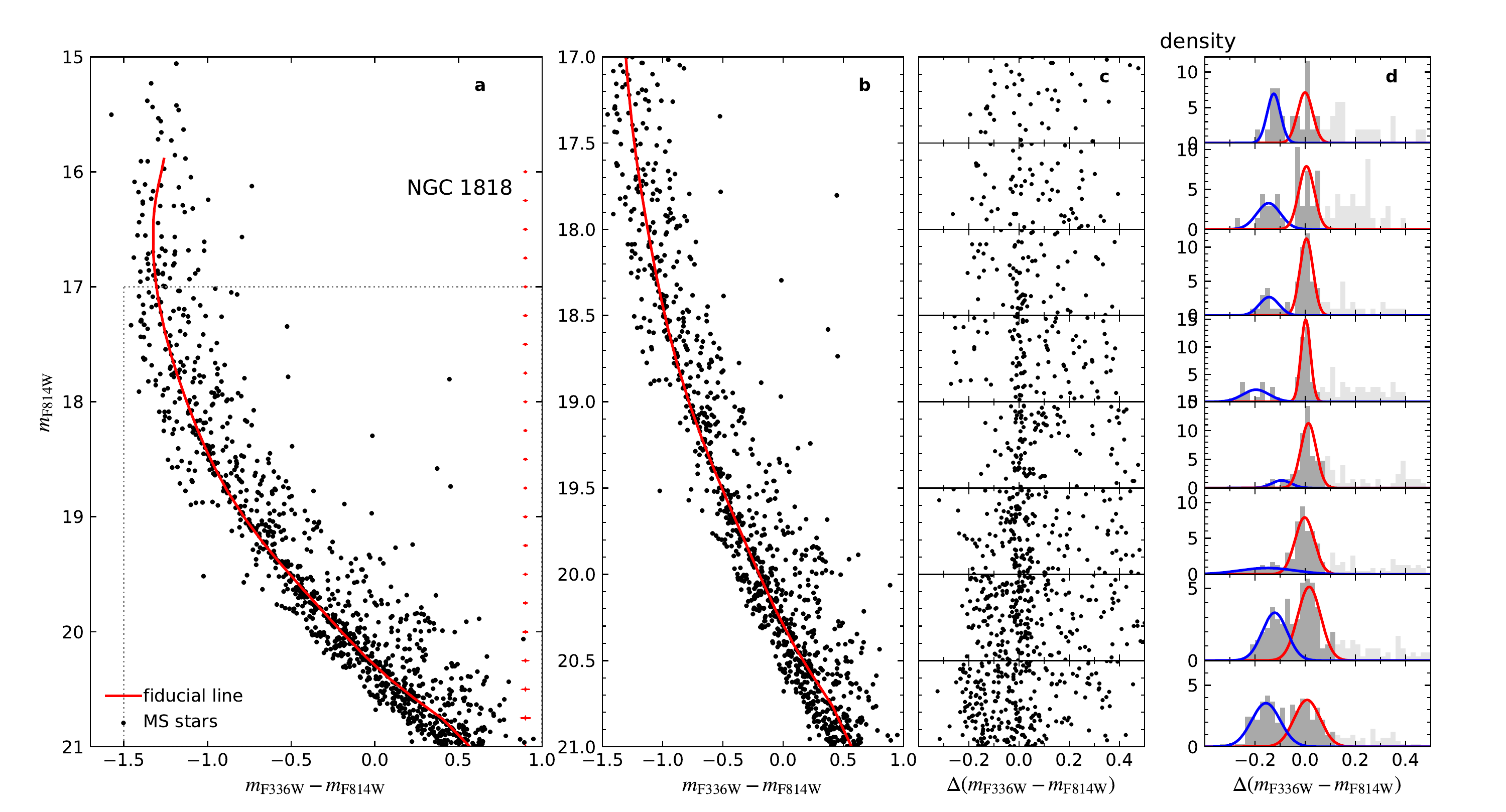}
\caption{Same as Supplementary Figure\,1, but for distinguishing the red and the blue main-sequence stars in NGC\,1818.}
\label{Suppfig:n1818_split}
\end{figure}    

\begin{figure}[ht]
\centering
\includegraphics[width=\linewidth]{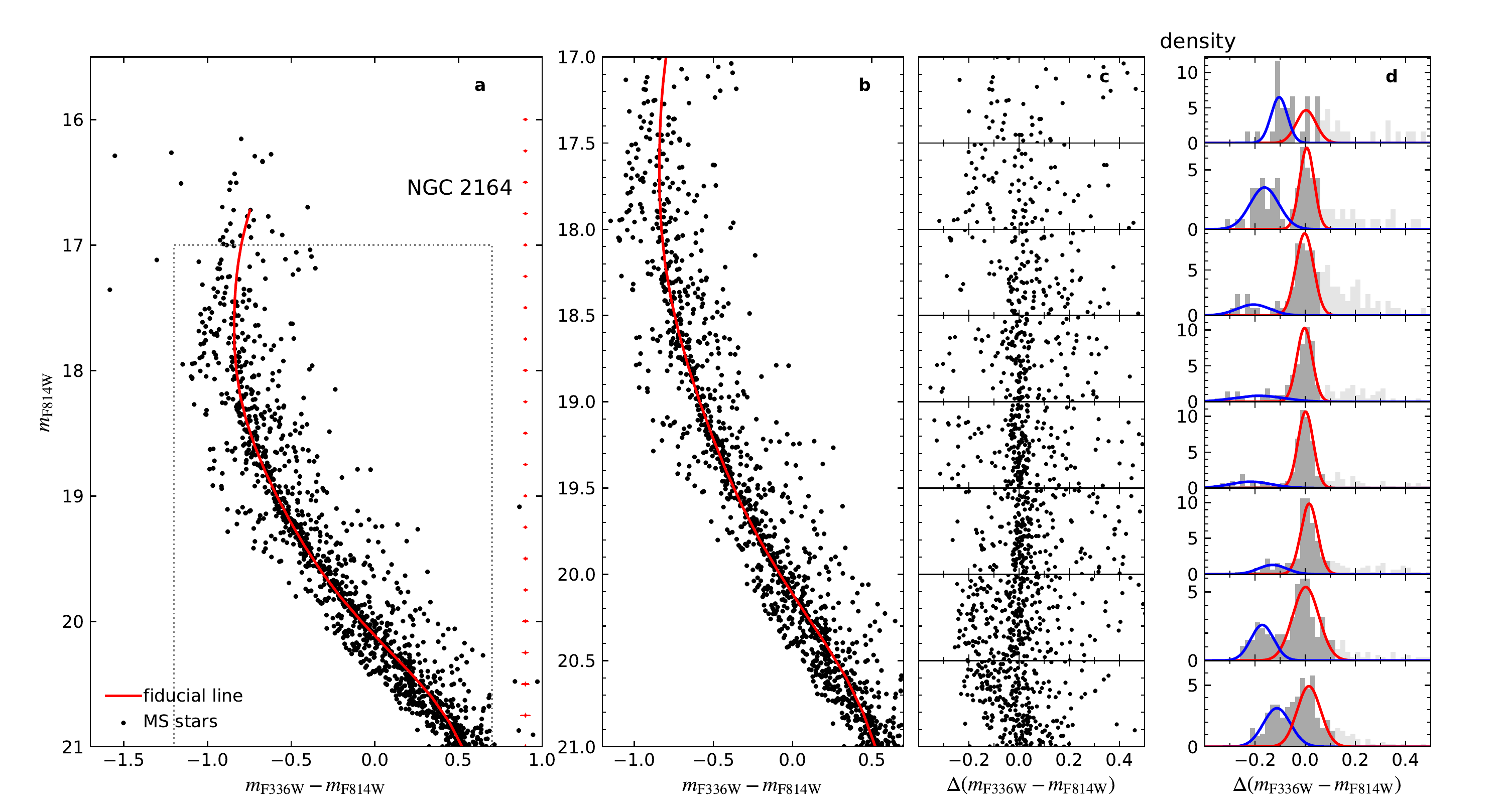}
\caption{Same as Supplementary Figure\,1, but for distinguishing the red and the blue main-sequence stars in NGC\,2164.}
\label{Suppfig:n2164_split}
\end{figure}    

\section*{B: The main sequence split as a function of stellar rotation}\label{suppsec:B}

Rotation is widely accepted to be responsible for the split MS \cite{Bastian+2009,2015MNRAS.453.2070N,2017MNRAS.467.3628C}. In this section, we explore how much rotation is required to retrieve the color split by comparing our rotating single star models with the MS components in the above mentioned four clusters. We next describe the physics and assumptions adopted while computing the single star models.

In the framework of a single star-burst forming a star cluster, (i.e., all stars are born at the same time), we attempt to use the isochrones constructed from our stellar models to fit the observations. The observations show a narrow red MS band, which marks the peak of stellar density in the CMD. 
In particular below 19th magnitude, the CMD of NGC\,1755 (see Fig.\,1a) shows a clear gap between the red MS and bluer stars. 
We start the isochrone fitting by adopting the parameters (isochrone age, distance modulus and reddening) derived in \cite{2018MNRAS.477.2640M}. Since we are using different
stellar models, we then need to adapt these parameters together with the stellar rotation parameter gradually until we obtain a pair of isochrones with the same age, but with different initial rotational velocities that can simultaneously best match the red MS and the blue MS bands, by visual inspection. We find that stellar models with $W_{\rm i}=0.65$ and $W_{\rm i}=0.35$ can fit the observed red and blue MSs equally well in all the clusters analyzed here (see Fig.\,1b and Supplementary Figure\,1). The adopted parameters in the isochrone fitting, as well as the mass of the studied clusters provided in \cite{2018MNRAS.477.2640M} are listed in Supplementary Table\,1. 
The adopted parameters in our work are slightly different from \cite{2018MNRAS.477.2640M}, as a consequence of different employed stellar models, with slightly different initial chemical composition and adopted physics parameters. The resulting small differences in the
fit parameters of the employed isochrone are not significant, 
because only the relative distance between red and blue isochrone well below the turn-off is important for our analysis.
The bluer isochrone identifies the bulk of blue MS stars to the blue of the red isochrone,
with, however, more and more stars falling to its blue side with higher brightness.

The stars to the red (right) side of the red MS
are most likely unresolved close binaries, which are expected to lie in between the red MS and the corresponding
equal-mass binary line, constructed by adding 0.75 magnitudes to the isochrone fitting the red MS, corresponding to a factor of two in flux (red dashed line in Fig.\,1b and Supplementary Figure\,5).  This interpretation is strengthened by a rather sharp drop in stellar density to the red side of the equal-mass binary line seen in all the analyzed clusters. 
Since the location of this drop coincides well with
the equal-mass binary isochrone derived from the red MS,
we expect that most of the unresolved binaries, as most stars in these clusters generally, are rapid rotators.
Redder stars are likely Be stars 
when near the turn-off \cite{2018MNRAS.477.2640M}, or higher order multiple systems otherwise.

We notice that the stars to the red (right) side of the red MS may also be explained by stars rotating with even faster velocities.
Supplementary Figure\,6 shows a comparison of isochrones derived from our single star models computed with different initial rotational rates, with the MS stars of NGC\,1755. 
The adopted age, distance modulus and reddening are the same as in Fig.\,1b. 
We see that the faster rotating stellar models are redder than the slower rotating ones, due to their lower effective gravity.
For slow rotation, an increase of the rotation parameter $W_{\rm i}$ by 0.1 has only a small effect on the isochrone color. However, it becomes progressively larger for faster rotation. 
The isochrone of our fastest rotating models ($W_{\rm i}=0.75$) overlaps the CMD region populated by suspected unresolved binaries (with large mass ratios).
However the majority of the stars redder than the red MS cannot be interpreted as extremely fast rotating stars, otherwise it contradicts the observed rotational velocity distributions of the B-type and A-type stars \cite{Huang+2010,2012A&A...537A.120Z,2013A&A...550A.109D} and the fact that H$\alpha$ emitters have only been detected in the region within two magnitudes below the turn-off \cite{2018MNRAS.477.2640M}.

Supplementary Figure\,6 can also be used to constrain the width of the rotational velocity distribution of the stars on the red MS.
Its broadening can be delineated well by single stars with $W_{\rm i}$ from 0.45 to 0.65.  Even though binaries composed of two slow rotators can also occupy the red MS, we do not expect them to play an important role if most slow rotators originate from binary mergers, after which most of them should be single stars. Nevertheless, low-mass ratio tidally-locked binaries may provide a small contribution to the red MS population (see Supplementary Information C and Supplementary Figure\,8).

The choice of the rotation parameter for the isochrone fitting of the red MS is slightly degenerate, such that slightly smaller rotation parameters may also provide
acceptable fits.
We investigate how a different choice of $W_{\rm i}$ would impact our conclusion for the four clusters in Supplementary Figure\,7. The adopted parameters are listed both in the figure and in Supplementary Table\,1. 
Even though rotation parameters of $W_{\rm i}=0.55$ and $W_{\rm i}=0.15$ can retrieve the observed color split of the red and blue MS equally well compared to the values of $W_{\rm i}=0.65$ and $W_{\rm i}=0.35$, the former lead to several stars being bluer than the zero-age MS line in NGC\,1755, NGC\,330 and NGC\,1818, which could not be interpreted by stellar rejuvenation caused by binary mergers.
Notably, the remaining degeneracy of the rotation parameter and age for the isochrone fitting does not affect our main conclusions, because it is not the precise values of these two parameters, but the gap in the CMD which determines our results.
E.g., the derived mass functions of the red and blue MS stars change only marginally for different isochrone fits (see Supplementary Information D and Supplementary Figure\,13). We show in Supplementary Information E that our conclusion of a high frequency of stellar merger events during the early cluster evolution also holds for different isochrone fits.

While it is not in the focus of this work, it is worth mentioning that the
bi-modal distribution of rotation rates as adopted here is well suited to lead to an extended 
main sequence turn-off as it is observed in most of the young open clusters. While binary evolution
is known to also contribute substantially \cite{2020ApJ...888L..12W}, an initial rotational velocity of around 50\% of critical is large enough to considerably widen the turn-off region (see Supplementary Information C and Fig.\,3). The reason is that MS stars increase the
ratio of rotation to critical rotation velocity during their evolution \cite{2008A&A...478..467E,2020A&A...633A.165H}, such that the extended main sequence turn-off will be significantly enhanced by the inclination dependence of gravity darkening.

\begin{figure}
 \centering
 \includegraphics[width=0.45\linewidth]{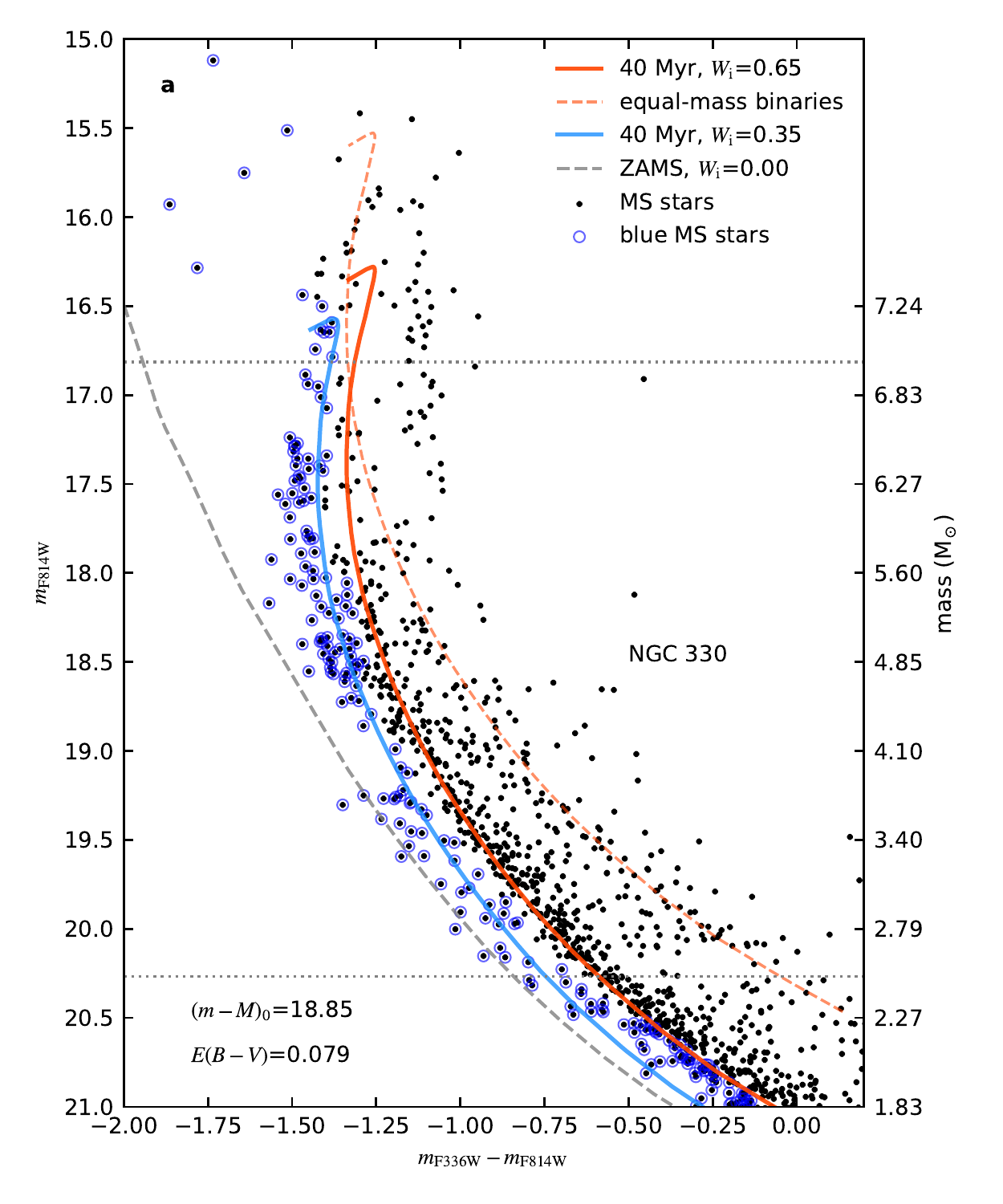}
 \includegraphics[width=0.45\linewidth]{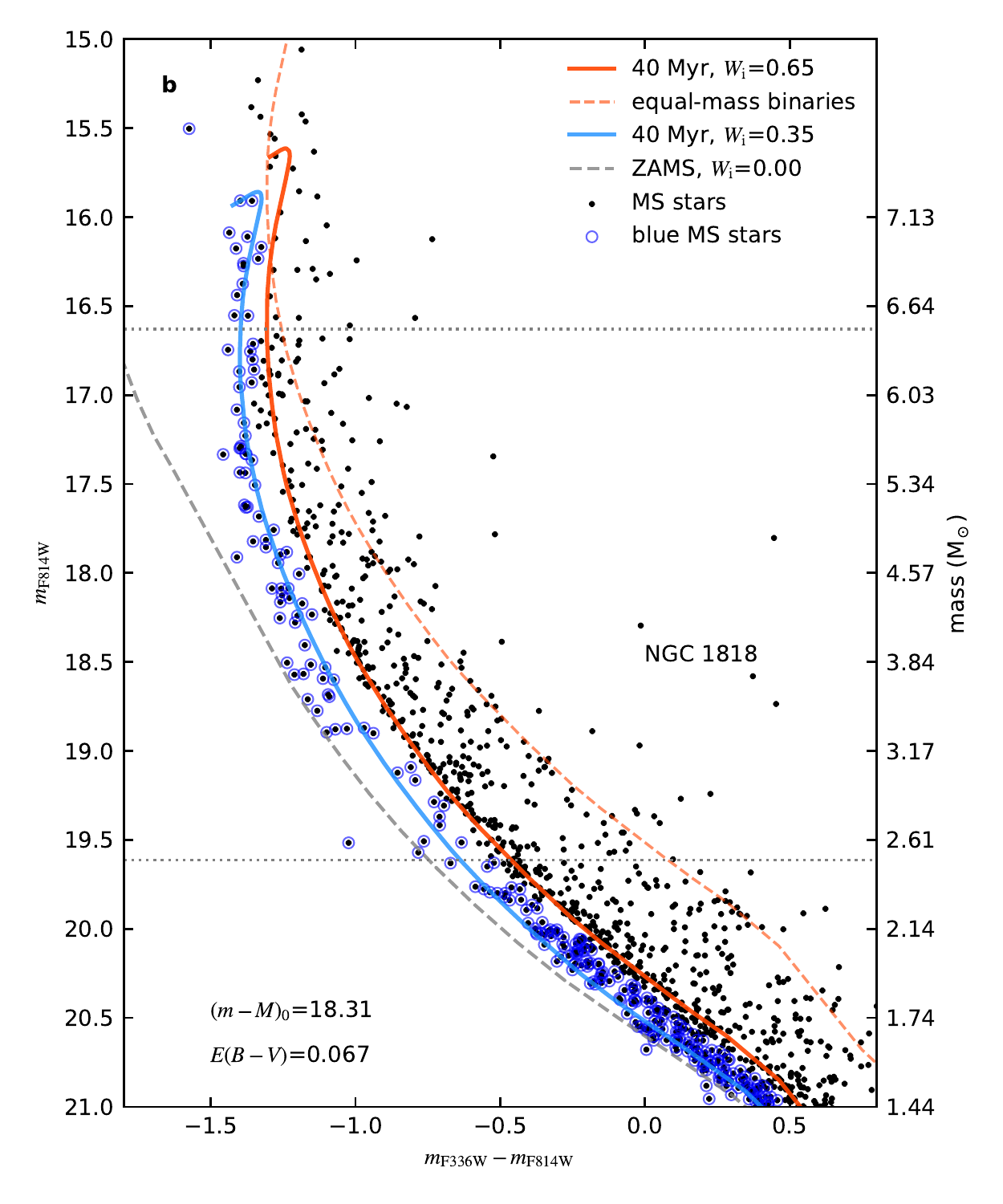}
 \includegraphics[width=0.45\linewidth]{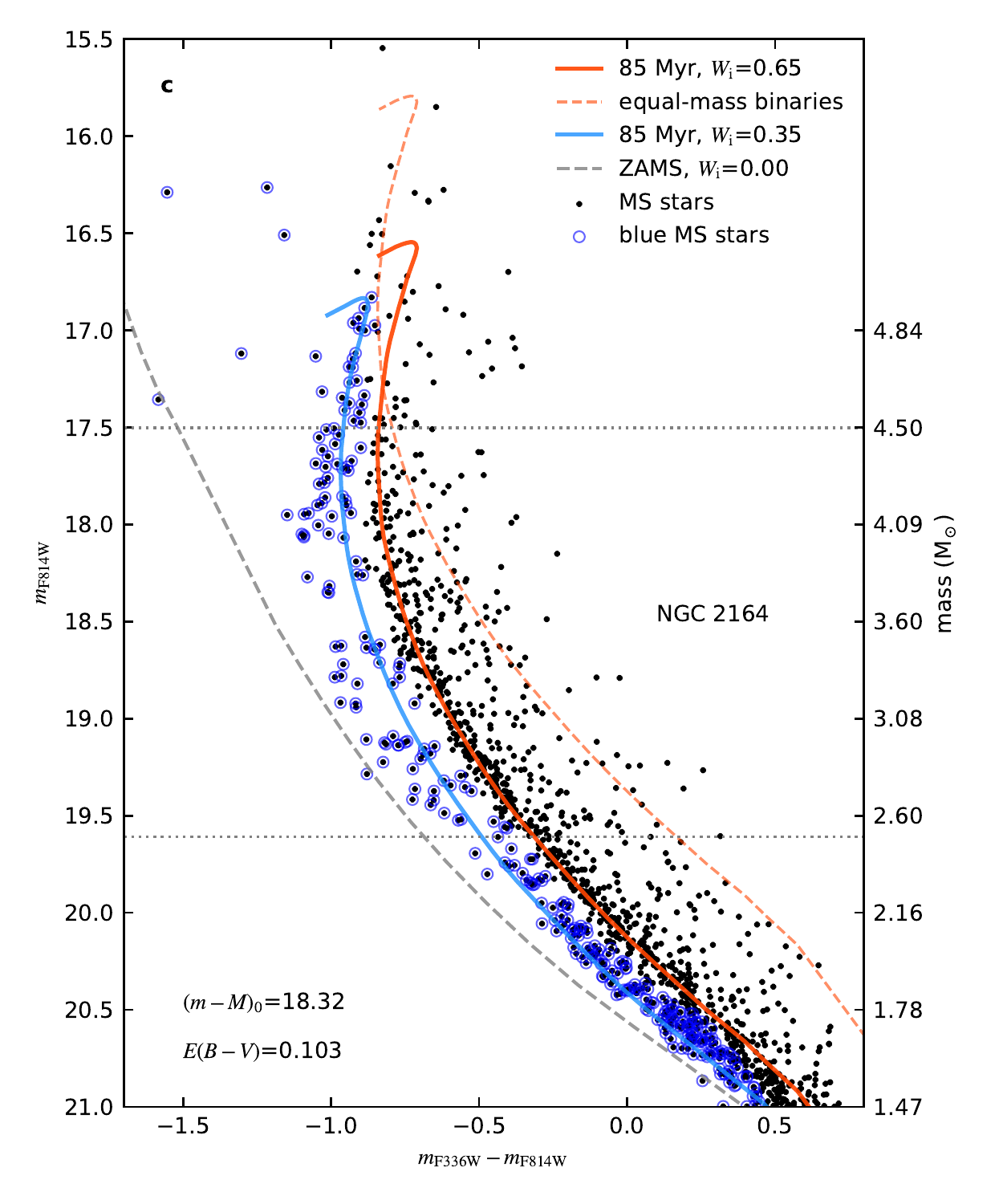}

 \caption{Isochrone fits to the main-sequence stars in three young star clusters. The plots are the same as Fig.\,1b. The adopted distance moduli and reddenings are indicated (see also Supplementary Table\,1).}
\label{Suppfig:3syn}
\end{figure}

\begin{figure}[ht]
\includegraphics[width=0.8\linewidth]{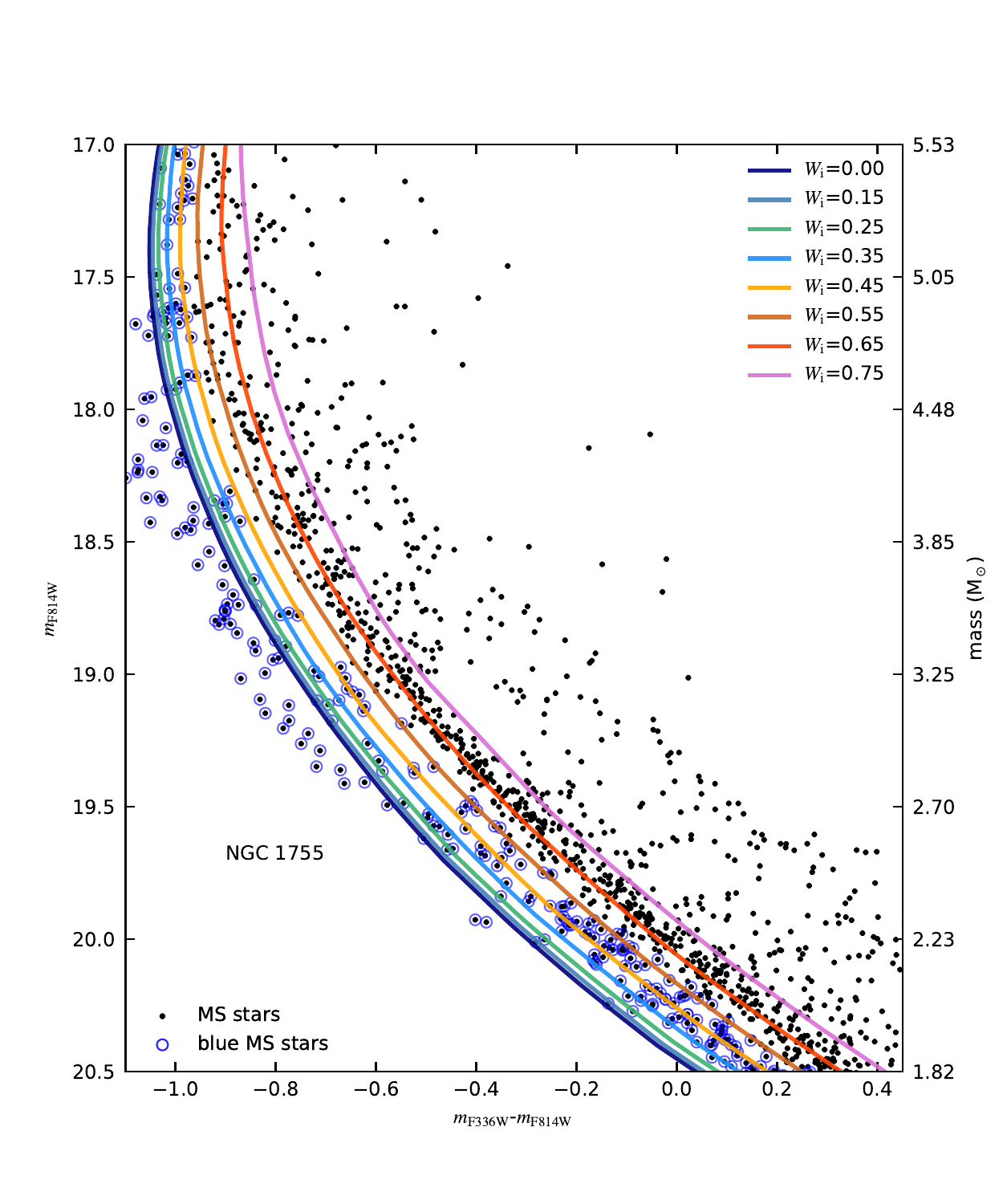}
\caption{Effect of rotation on the color of main-sequence stars. The black dots and blue open circles correspond to the observed main-sequence stars and the identified blue main-sequence stars in NGC\,1755, respectively. The isochrones are derived from our single star models with different initial rotational rates,
as indicated in the legend. The adopted distance modulus and reddening are the same as Fig.\,1b, as are the isochrones with a rotation
parameter of $W_{\rm i}=0.35$ and $W_{\rm i}=0.65$.  
The right y-axis displays the stellar masses derived from the mass-magnitude relation of the models with 65\% of critical rotation initially.
}
\label{Suppfig:n1755_rotation}
\end{figure}

\begin{figure}[ht]
\centering
\includegraphics[width=0.4\linewidth]{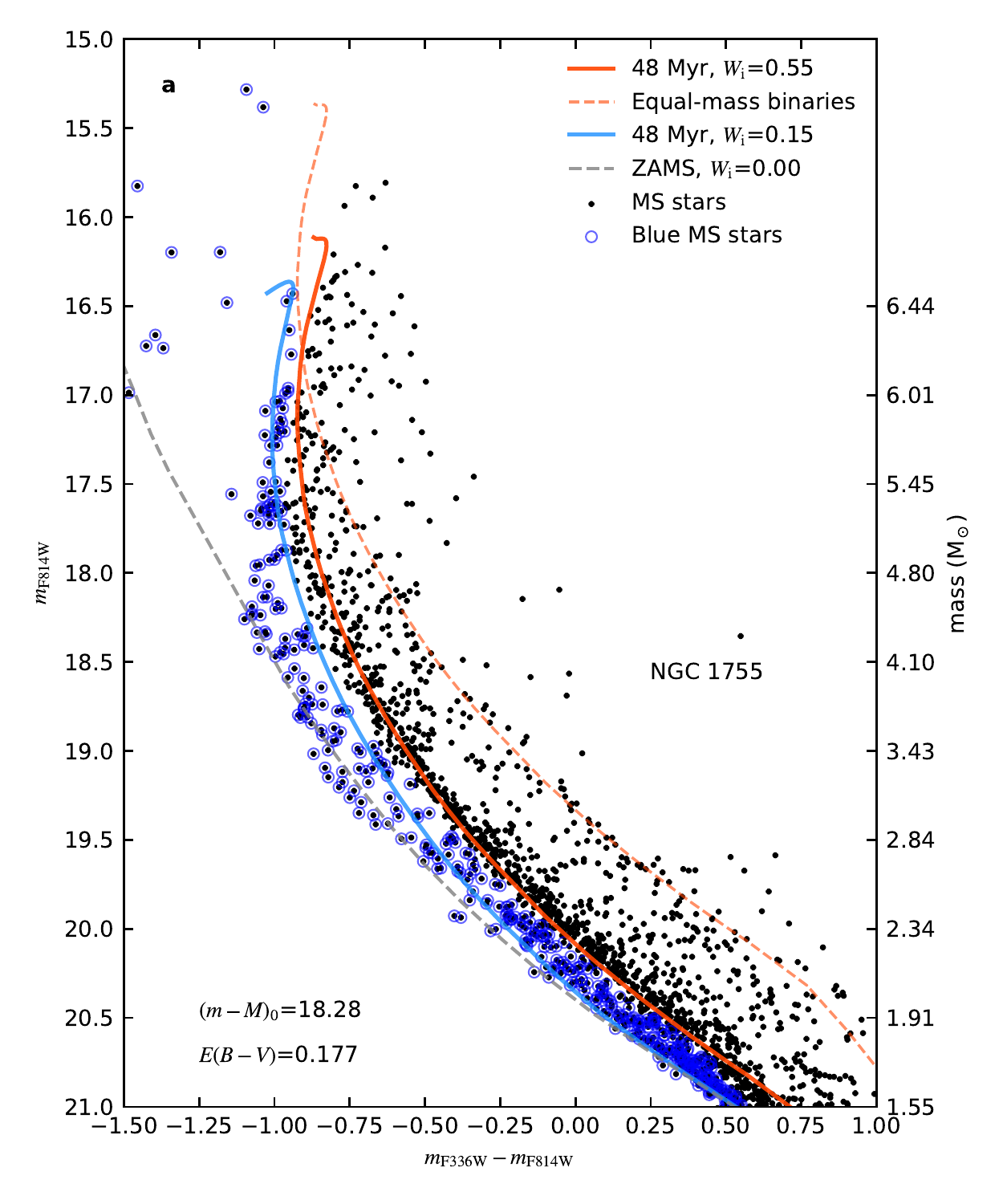}
\includegraphics[width=0.4\linewidth]{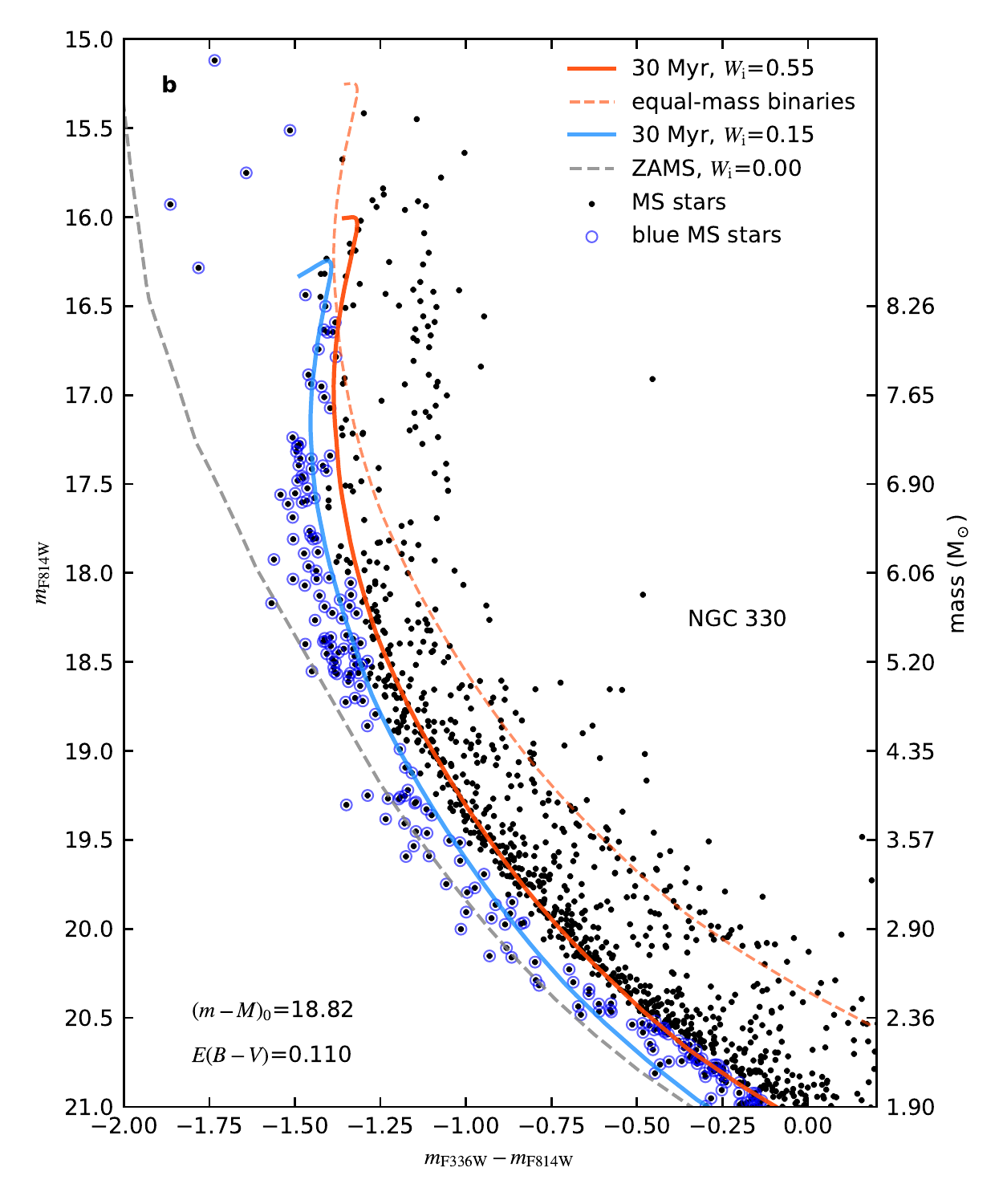}
\includegraphics[width=0.4\linewidth]{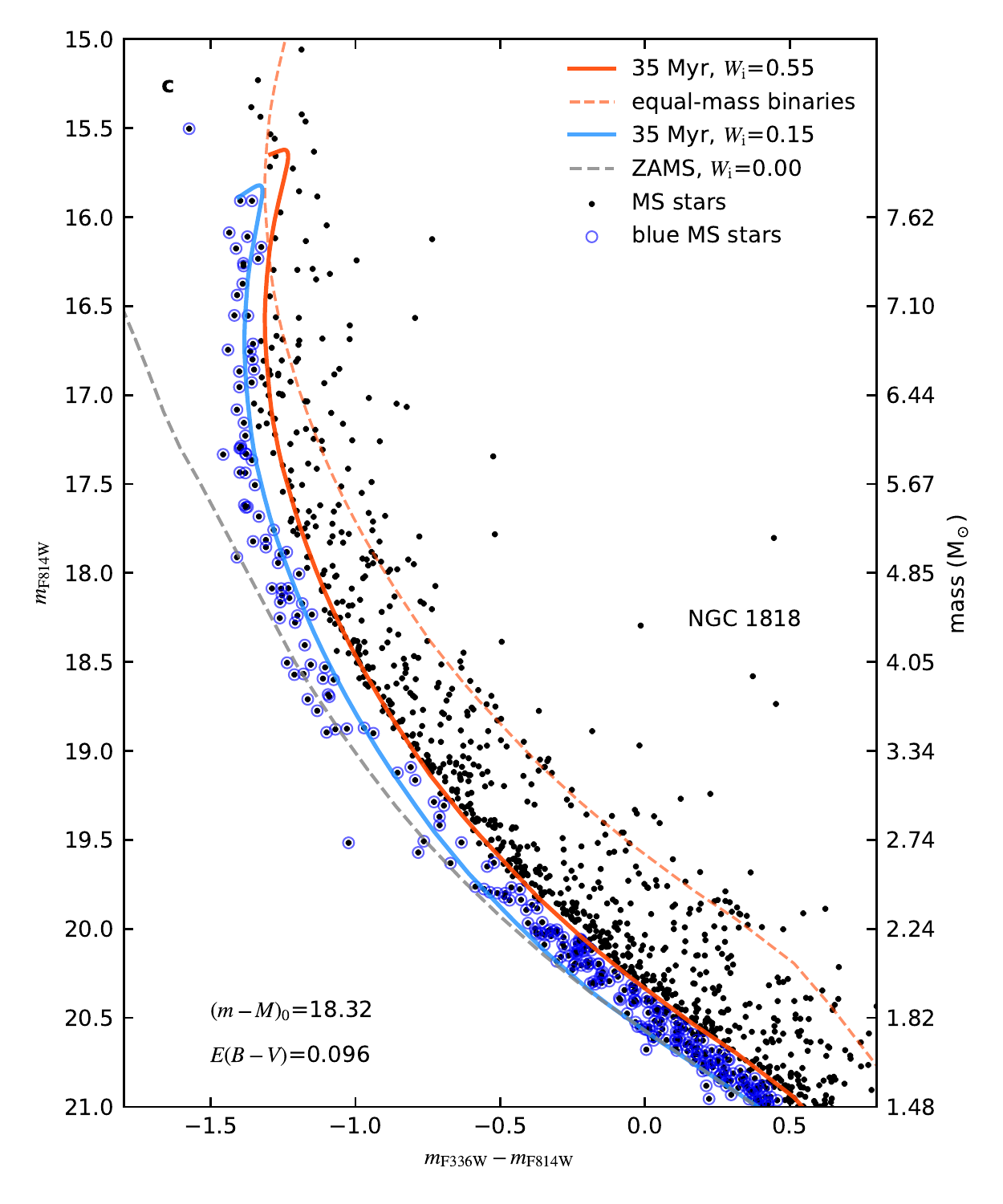}
\includegraphics[width=0.4\linewidth]{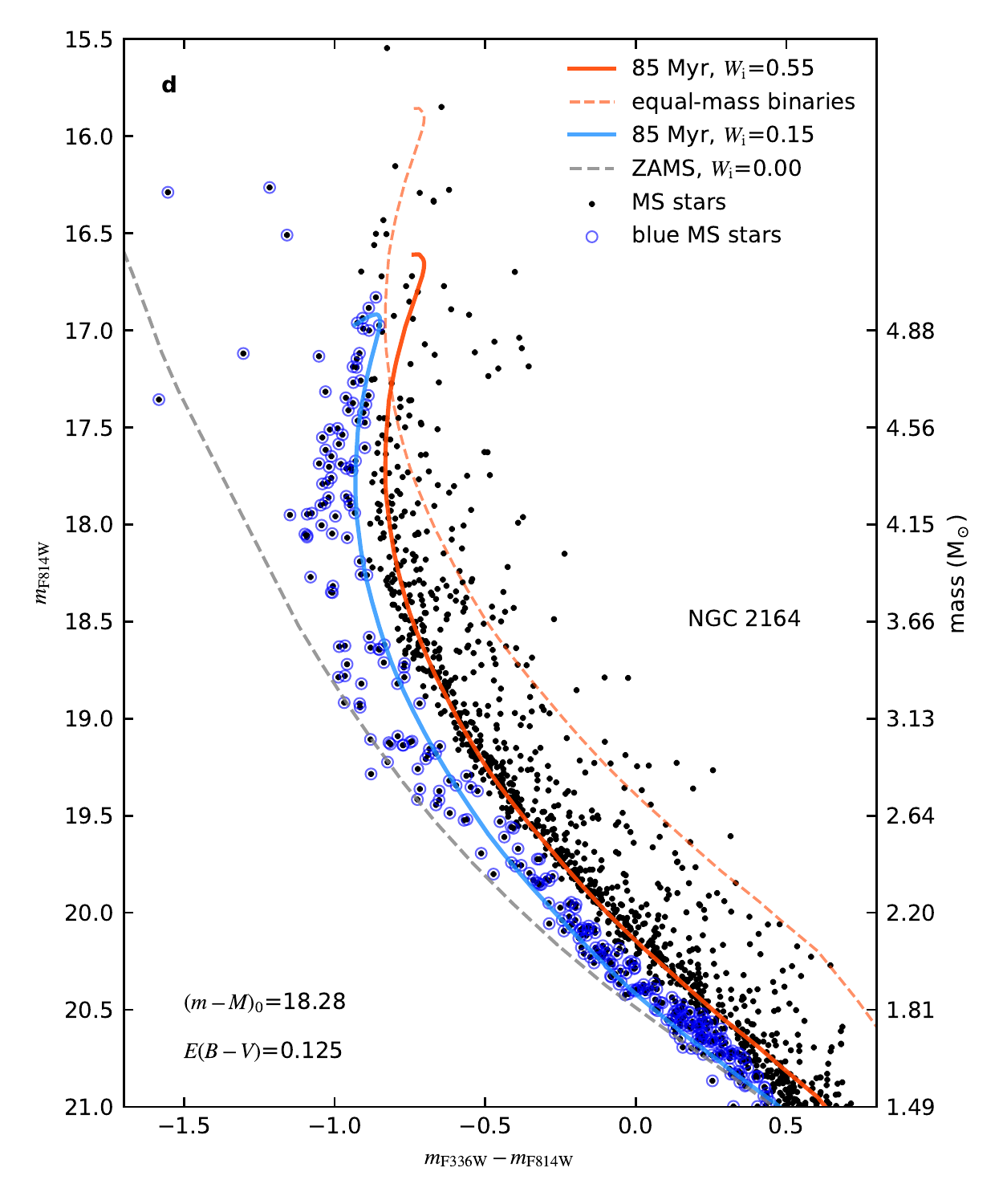}

\caption{Isochrones for four young star clusters using alternative stellar rotation. The plots are the same as Fig.\,1b, but stellar models with different initial rotational velocities ($W_{\rm i}=0.55$ and $W_{\rm i}=0.15$) are employed to fit the red and blue main-sequences of the clusters (see legends). The adopted distance moduli and reddenings are indicated both in the figure and in Supplementary Table\,1. 
}
\label{Suppfig:4syn_v55}
\end{figure}         

\setcounter{table}{0}
\renewcommand{\tablename}{Supplementary Table}

\begin{table}
\begin{center}
\caption{Basic information of the studied clusters and the parameters adopted in fitting the observed red and blue main sequences with our single star models. The information of the cluster mass is from \cite{2018MNRAS.477.2640M}.}
\scalebox{0.9}{
\begin{tabular}{ l c c c c c c c} 
\toprule
\midrule
Cluster  & Galaxy & log$(M/M_\odot)$& $W_{\mathrm i}$ for red MS & $W_{\mathrm i}$ for blue MS & Age (Myr)  &  $(m-M)_0$  & $E(B-V)$ \\
\midrule
NGC\,330    &  SMC   & 4.61 & 0.65  & 0.35   & 40    & 18.85   & 0.079 \\
NGC\,1818  &  LMC   & 4.41 & 0.65  & 0.35   & 40    & 18.31   & 0.067 \\
NGC\,1755  &  LMC   & 3.60 & 0.65  & 0.35   & 58    & 18.29  & 0.140 \\
NGC\,2164  & LMC    & 4.18 & 0.65  & 0.35   & 85    &18.32    & 0.103 \\
\midrule
NGC\,330    &  SMC   & 4.61 & 0.55  & 0.15   & 30    & 18.82   & 0.110 \\
NGC\,1818  &  LMC   & 4.41 & 0.55  & 0.15   & 35    & 18.32   & 0.096 \\
NGC\,1755  &  LMC   & 3.60 & 0.55  & 0.15   & 48    & 18.28   & 0.177 \\
NGC\,2164  & LMC    &  4.18 & 0.55  & 0.15   & 85    & 18.28    & 0.125 \\
\bottomrule
\end{tabular}
}
\end{center}
\end{table}    

\section*{C: Distribution of the detailed binary models in the CMD}\label{suppsec:C}
In the main text, we have proposed that single and binary stars in young star clusters are born with nearly the same velocities with values slightly larger than half of their break-up velocities. In this section, we examine the distribution of our detailed binary models in the CMD, attempting to inspect the contribution of binary evolution to blue MS stars. 

In Fig.\,3, we show the CMD distribution of our binary models and binary-evolutionary products at 30\,Myr. 
The magnitude of each binary model is obtained by adding the fluxes of the two components in the corresponding filter bands. The effect of gravity darkening is implemented according to \cite{2011A&A...533A..43E,Paxton2019} by assuming a random orientation for the rotational axis of a star model. We assume that the orbit and spin vectors have the same orientation. We assign an additional shift to each binary model (or binary merger product) in the CMD by considering a Gaussian distribution for the photometric errors at the corresponding magnitudes. For comparison, we overplot the observed MS stars in NGC\,330. 
We do not normalize the total number of our binary models to the observed number of stars, because a concrete quantitative comparison between our binary models and the observations is beyond the scope of this work. 

In agreement with \cite{2020ApJ...888L..12W}, we see in Fig.\,3 that MS mergers produce a population of blue stragglers on the left side of the turn-off, between the zero-age MS line and the solid blue line. 
We find a deficit of MS merger products fainter than $\sim$18.5\,mags at this age, because the faint stars far below the turn-off hardly have time to expand and undergo mass transfer. Besides, our models predict very few MS merger products near the solid blue line, due to the same reason. Therefore, orbit decay during the early evolution of the binaries (see the main text) is mandatory to explain the observed dense distribution of the blue MS stars near the solid blue line.
The modelling of this orbit decay is beyond the capabilities of current stellar evolution calculations.

Similar to the results in \cite{2020ApJ...888L..12W}, our newly computed binary models predict a sequence of critically-rotating stars to the red side of the turn-off region. These are the mass gainers of Case B mass transfer, which reach critical rotation and avoid tidal spin down, and likely correspond to Be stars. To account for the
flux contribution from the decretion disk of these stars,
we increase their red magnitude by 0.2\,mags \cite{2017AJ....153..252L,2021arXiv210612263H}.

In our shortest period binary evolution models, the rotation of the two components is affected by tides after the zero-age MS. 
Fig.\,3 shows a population of tidally-braked binary models, in which the two stars rotate at velocities lower than their initial values.
Such systems, if they have small mass ratios, 
may be located in positions in the CMD that we assign to hold blue MS stars. Additionally, some of our binary models contain a MS star and a stripped, hot helium burning star, which can also contribute to the observed blue MS stars. 
To explore the degree of contamination of blue MS from these binaries, we calculate the fraction of our detailed binary models (MS$+$MS binaries and MS$+$He burning star binaries) which would be classified as blue MS stars, with respect to the total number of binary models. 
We consider models as blue MS stars if their colors are bluer than the median color of the two isochrones shown in Fig.\,3 at corresponding magnitudes.
We do this calculation for our detailed binary models for ages from 15 to 100\,Myr. The result is shown in Supplementary Figure\,8.

We perform this analysis only for binary models which are at least one magnitude below the cluster turn-off, since blue and red MSs become
undistinguishable near the turn-off (Fig.\,3). This happens because the isochrone in this region is almost vertical, and as a consequence, the unresolved binaries (with whatever rotation) may be bluer than their single counterparts and contribute to the blue MS stars. In addition, the effect of gravity darkening is more pronounced near the turn-off, resulting in a large spread of the fast-rotating stars.

In Supplementary Figure\,8, we see that for all considered ages, the fraction of binaries that contribute to blue MS stars remains below 3\%, which is small compared to the fraction of the observed blue MS stars ($\sim$20\%).
Also Fig.\,3 shows that the majority of the tidally-braked binaries are still found above the red MS, due to the presence of a companion, despite their slow rotation. 

This result is consistent with an analysis of the initial binary parameters. According to our initial period distribution, we find that only 4.6\% of the binaries hold two stars rotating at velocities lower than 15\% of their critical velocities due to tides. We find that, in general, these slowly-rotating binaries can be classified as blue MS stars only if their initial mass ratios are smaller than $0.25$. Assuming a flat mass ratio distribution, we conclude that only $\sim$0.77\%  of the MS binaries can contaminate the blue MS stars.
This simple analysis agrees well with the results in Supplementary Figure\,8. Therefore, we conclude that the contribution of tidally-braked binaries to the observed blue MS stars is marginal.

\begin{figure}[ht]
\centering
\includegraphics[width=0.9\linewidth]{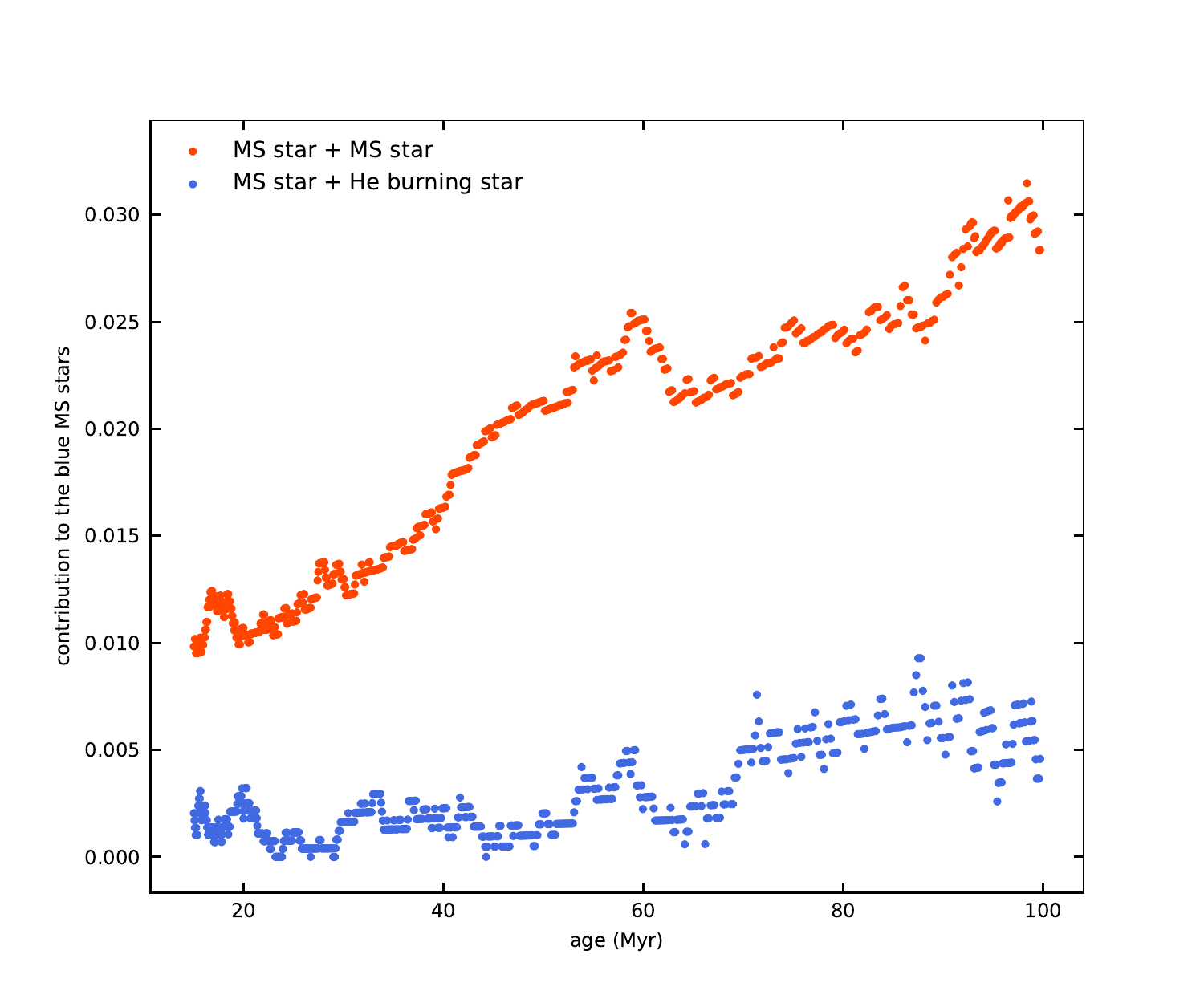}

\caption{Contribution of binaries (non-mergers) to the blue main-sequence for different ages. Red and blue dots correspond to the fraction of binary models containing two main-sequence stars and binary models containing a main-sequence star and a He burning star, that will be classified as blue main-sequence stars, with respect to all the binary models, respectively.
}
\label{Suppfig:binary_contamination}
\end{figure}     
\section*{D: Mass functions}\label{suppsec:D}

In this section, we investigate the mass functions for the red and blue MS stars in the above mentioned young star clusters. 
We mainly employ stars between the two grey dashed, horizontal lines in the CMD figures (Fig.\,1b, Supplementary Figure\,5), because the red and the blue MS cannot be well distinguished either above the upper grey dashed line due to the complexity of the turn-off stars, or below the lower grey dashed line due to the large photometric errors. The blue MS stars are those marked by blue circles, while the red MS stars are those not classified as blue MS stars. 

Taking NGC\,1755 as an example, we plot the cumulative number distribution of the identified red and blue stars in the mass range of $5.5\dots 2.5\mso$ with the solid red and blue lines in Fig.\,2a and Fig.\,2c.
We use the mass-magnitude relation contained in the isochrones of the $W_{\rm i}=0.35$ and $W_{\rm i}=0.65$ single star models to convert the magnitude of the identified blue and red MS stars to mass. 
We assume that stars in each population obey a power-law mass function $N(m)\,{\rm d}m \propto m^\gamma$,  where $m$ means the mass and $N(m)\,\rm{d}m$ means the number of stars with masses in the range $m$ to $\rm{d}m$. We then find the $\gamma$ value that can best fit the cumulative distribution of the observed stars in each population. We do the fitting in the cumulative distribution plane to avoid the uncertainties introduced by mass bins. The results are shown with color dashed lines in Fig.\,2, with the values of the mass function slopes and one sigma errors listed. One sigma error is calculated such that 68.3\% of the observed distribution (solid color line) is covered, shown by the colored shaded area. The residuals that describe the difference between the observed distribution and the predicted distribution are shown in Fig.\,2b and Fig.\,2d for the red and the blue MS stars, respectively.
We do the same for the red and the blue MS stars in the mass range of $1.8\mso$ to $2.5\mso$ in NGC\,1755 and show the results in Supplementary Figure\,9. 
It is clearly seen that even though the mass distribution of the blue MS stars smaller than $\sim 2.5 \mso$ is steeper than the Salpeter IMF and the mass function of the red MS stars in the same mass range, the distribution of the higher mass blue MS stars ($>2.5\mso)$ is significantly flatter. The shallow slope for the massive blue MS stars may relate to the fact that the binary fraction is larger for more massive stars. Additionally, a shallow slope may also result from binary orbital decay simulations suggesting that binaries with higher masses are more likely to merge than binaries with lower masses \cite{2012A&A...543A.126K,2020MNRAS.491.5158T}.

The derived mass function slope depends on the considered mass or magnitude range. We notice that in three of the four analysed clusters (NGC\,1755, NGC\,1818 and NGC\,2164), the mass functions of the blue MS stars below $\sim 2.5\mso$ are steeper than a Salpeter IMF. At the same time, the split MS persists until a brightness corresponding to stars of $\sim 1.5\mso$.
Nevertheless, we need to emphasise here that the distinction between the red and the blue MS stars below $\sim 2.5\mso$ becomes vague, which can be seen in Supplementary Figure\,1d that the common area of the two Gaussian components is large at $m_\mathrm{F814W}\geq 20$. Therefore, the derived mass function for the stars less massive than 2.5$\mso$ is less trustworthy than that of the more massive stars. We do not consider the stars fainter than a brightness corresponding to $\sim 2.5\mso$ in the SMC cluster NGC\,330, as the distance modulus of this cluster is larger than the three LMC clusters.

For our main result, we consider masses above $2.5 \mso$.
Interestingly, \cite{2012A&A...537A.120Z} found that in galactic
field MS stars below $\sim 2.5\mso$, the slowest rotators
have a significantly larger spin than above $\sim 2.5\mso$,
consistent with the redder color of the blue MS stars below $\sim 2.5\mso$ in the clusters considered here.
These findings may imply an intrinsic difference in star formation of the slowly-rotating stars above and below $\sim 2.5\mso$, perhaps caused by a mass dependent pre-MS evolution.
Due to the lack of spectroscopic observations and the larger photometric errors for the stars below $2.5\mso$, we do not include them in our analysis. Nevertheless, the mass functions of the blue MS stars above $\sim 2.5\mso$ undoubtedly reveal a discrepancy with the Salpeter IMF, i.e. our conclusion that the red and blue MS stars have different mass functions is robust for stars whose masses are larger than $\sim 2.5\mso$.

We do the same experiment for the SMC cluster NGC\,330 and the LMC clusters NGC\,1818 and NGC\,2164. 
The results are shown in Supplementary Figure\,10, Supplementary Figure\,11, and Supplementary Figure\,12, respectively.
We summarize the derived slopes and errors of the mass functions of the MS stars more massive than $2.5\mso$ in different clusters in Supplementary Table\,2. Even though the error of the derived slope of the blue MS stars in NGC\,2164 is large, the slope distinction between the red and blue MS stars clearly exists. 
We found no significant differences between the mass function slopes measured across the four clusters, thus we argue that the mass function dichotomy is ubiquitous in all the studied clusters.

To assess to what extent the derived $\gamma$ values are affected by blue MS star classification, we examine different boundary lines to define the red and the blue MS stars, based on isochrone fitting.
We consider boundary lines to lie on the right side of the isochrone for the blue MS, with a color difference equal to a fraction $Q$ of the color separation between the isochrones for the red and the blue MSs. $Q$ ranges from 0 to 1, in intervals of 0.2, with $Q=0$ ($Q=1$) representing the extreme case that all stars on the left side of the isochrone for the blue (red) MS are assigned as blue MS stars. The stars that are not classified as blue MS stars are considered as red MS stars. We only consider the stars between the two grey dotted lines in Fig.\,1 and Supplementary Figure\,5. The variation of the derived $\gamma$ values for the red and the blue MS stars with respect to $Q$ in the four clusters is shown in Supplementary Figure\,13. 
It reveals that the mass functions of the red MS stars more massive than $2.5\mso$ in all four clusters are comparable to the Salpeter IMF. While the mass functions of the blue MS stars in the same mass range in all four clusters have significantly shallower slope than the Salpeter IMF, with the slope becoming slightly steeper as we use redder boundary lines. Nevertheless, the disparity between the mass function of the blue MS stars and the Salpeter IMF exists even in an extreme case of $Q=1$.

At last, we use the same stars as we employed in mass function estimation to calculate the fraction of the blue MS stars. The results are shown in Supplementary Table\,2, with the mean values obtained by assuming $Q=0.5$ when computing the borderline between the red and the blue MS stars, and the lower and upper errors obtained by assuming $Q=0$ and $Q=1$, respectively. We found that the ratio of the blue MS stars is almost identical in all four clusters.

\begin{table}
\begin{center}
\caption{Slopes $\gamma$ and uncertainties of the mass functions derived for the red and blue main-sequence stars with estimated masses larger than $2.5\mso$ in four clusters. The last column shows the number ratio between the blue main-sequence stars and all stars in the same mass range.}
\scalebox{0.9}{
\begin{tabular}{ l c c c c} 
\toprule
\midrule
Cluster  & Galaxy & $\gamma$ for the red MS stars & $\gamma$ for the blue MS stars  & $N_{\rm blue\, MS}/(N_{\rm blue\, MS}+N_{\rm red\, MS})$\\
\midrule
NGC\,330    &  SMC   & $-2.37 \pm 0.28$  & $0.37 \pm 0.39$    & $0.19^{+0.18}_{-0.07}$\\
NGC\,1818  &  LMC   & $-1.90 \pm 0.20$  & $-0.13 \pm 0.31$   & $0.15^{+0.13}_{-0.05}$\\
NGC\,1755  &  LMC   & $-2.17 \pm 0.15$  & $-1.03 \pm 0.32$   & $0.18^{+0.21}_{-0.06}$\\
NGC\,2164  & LMC    & $-2.38 \pm 0.43$  & $0.04 \pm 1.20$  & $0.15^{+0.23}_{-0.05}$\\
\bottomrule
\end{tabular}
}
\end{center}
\end{table}

\newpage

\begin{figure}[ht]
\centering
\includegraphics[width=\linewidth]{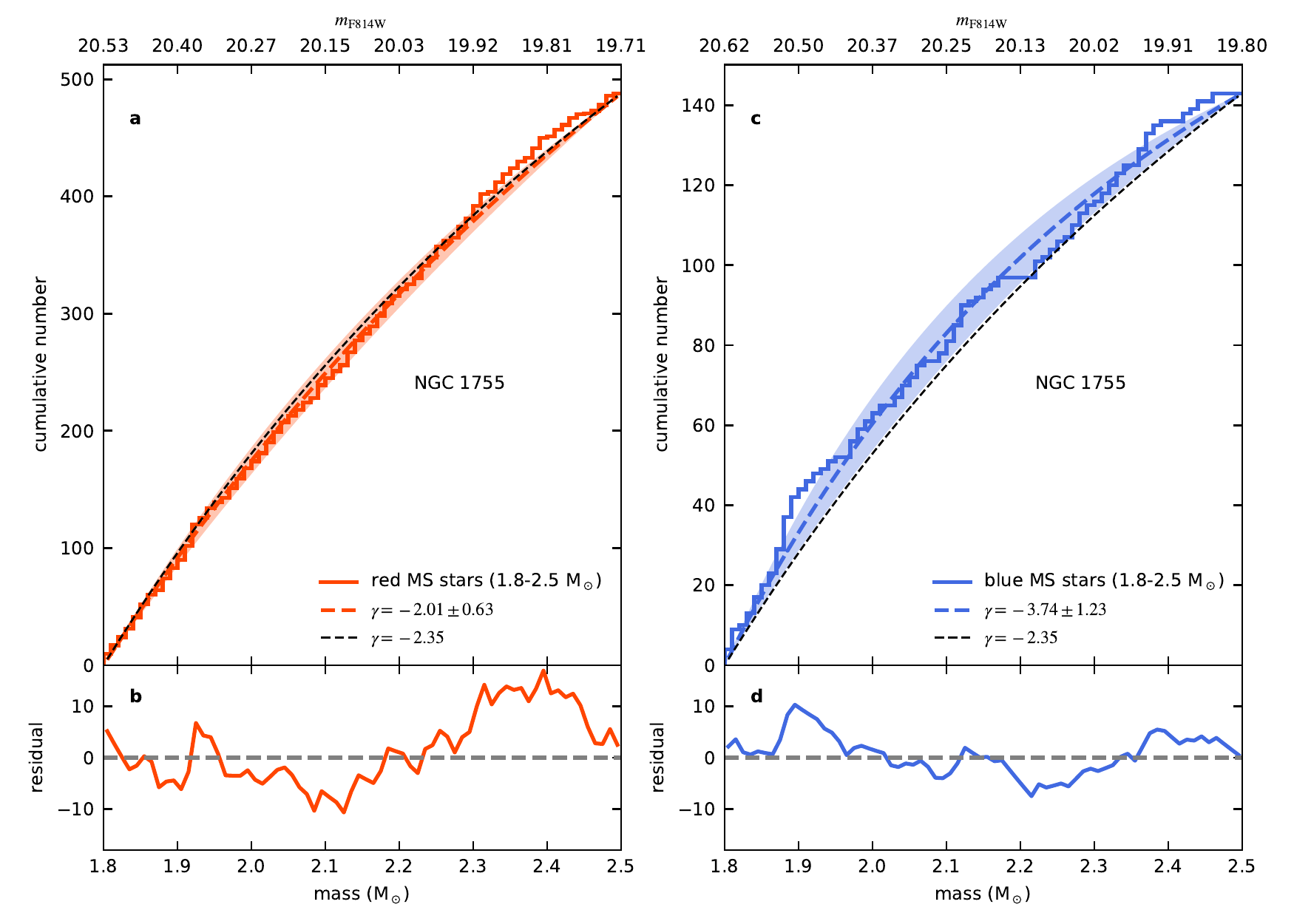}
\caption{Mass function of the low-mass red and blue main-sequence stars in NGC\,1755. The plots are the same as Fig.\,2. The considered mass range is $1.8$ to $2.5\mso$.
}
\label{fig:n1755_MF2}
\end{figure}

\begin{figure}[ht]
\centering
\includegraphics[width=\linewidth]{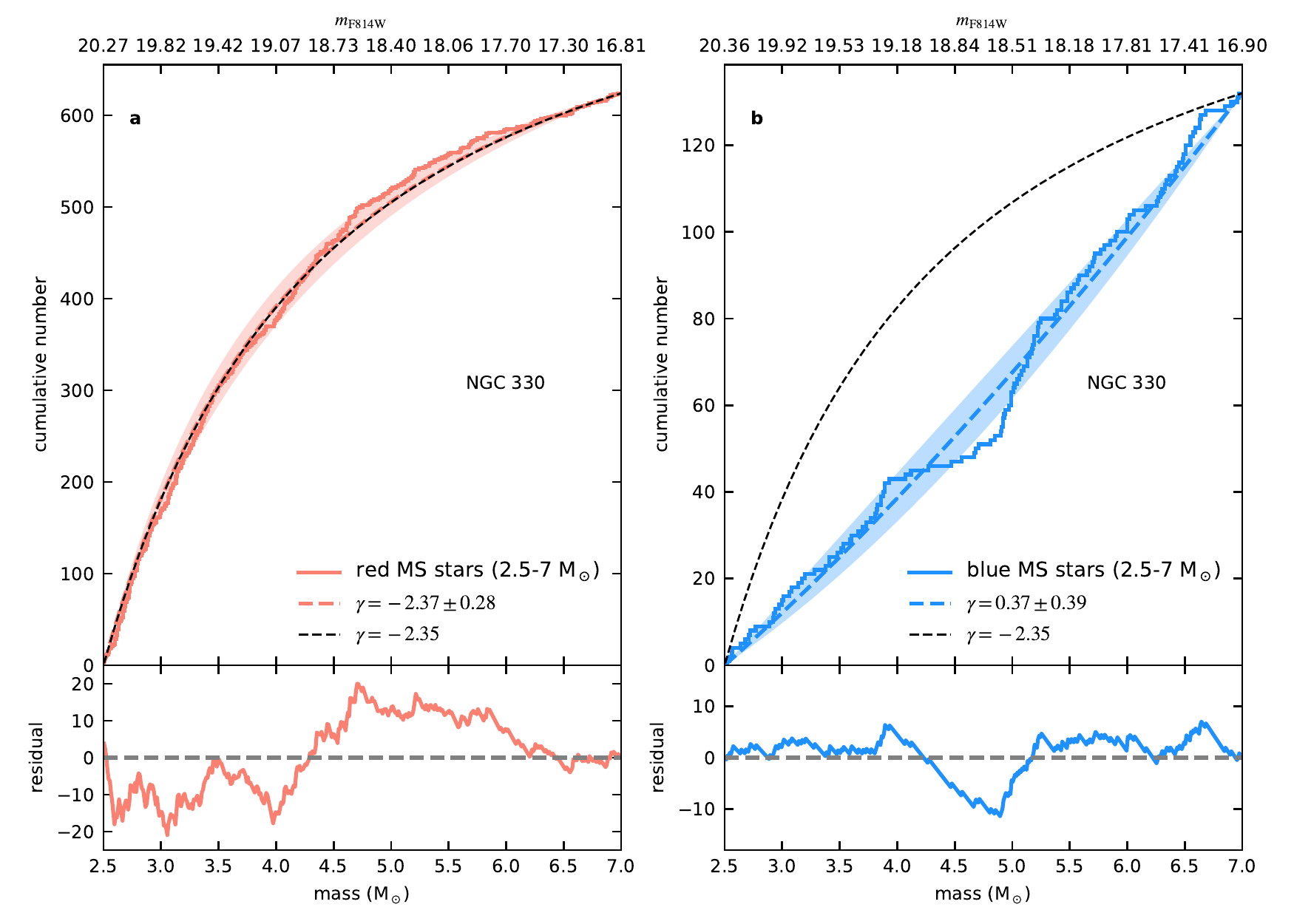}
\caption{Mass function of the red and the blue main-sequence stars in NGC\,330. The plots are the same as Fig.\,2. Only the stars with derived masses larger than $2.5\mso$ are considered.
}
\label{Suppfig:n330_MF}
\end{figure}

\begin{figure}[ht]
\centering
\includegraphics[width=\linewidth]{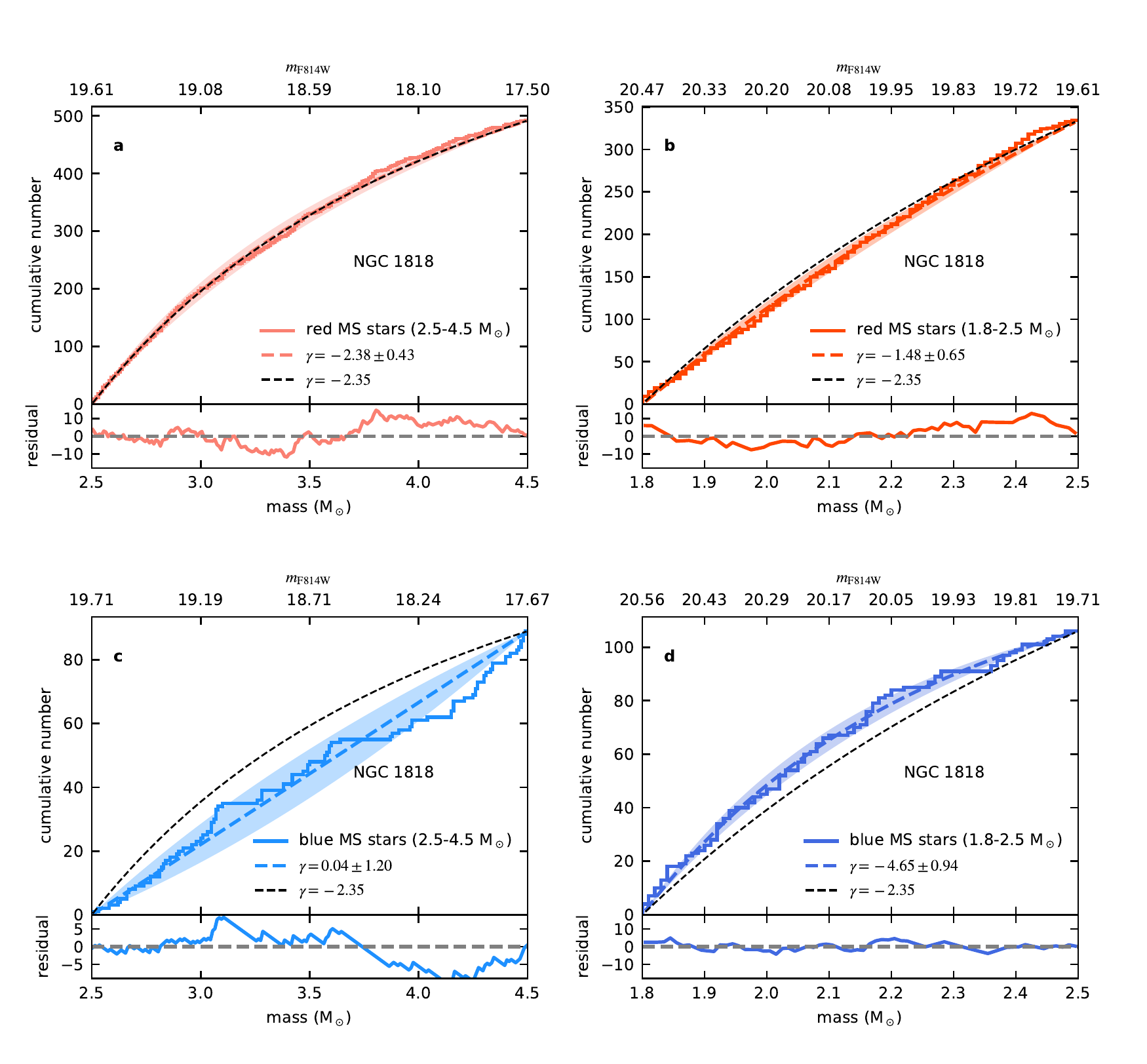}
\caption{Mass function of the red and the blue main-sequence stars in NGC\,1818. Panels a and c correspond to the results of the red and the blue MS stars in the mass range of $2.5$ to $6.5\mso$, respectively. While Panels b and d show the results of the red and the blue MS stars in the mass range of $1.8$ to $2.5\mso$, respectively.
}
\label{Suppfig:n1818_MF}
\end{figure}

\begin{figure}[ht]
\centering
\includegraphics[width=\linewidth]{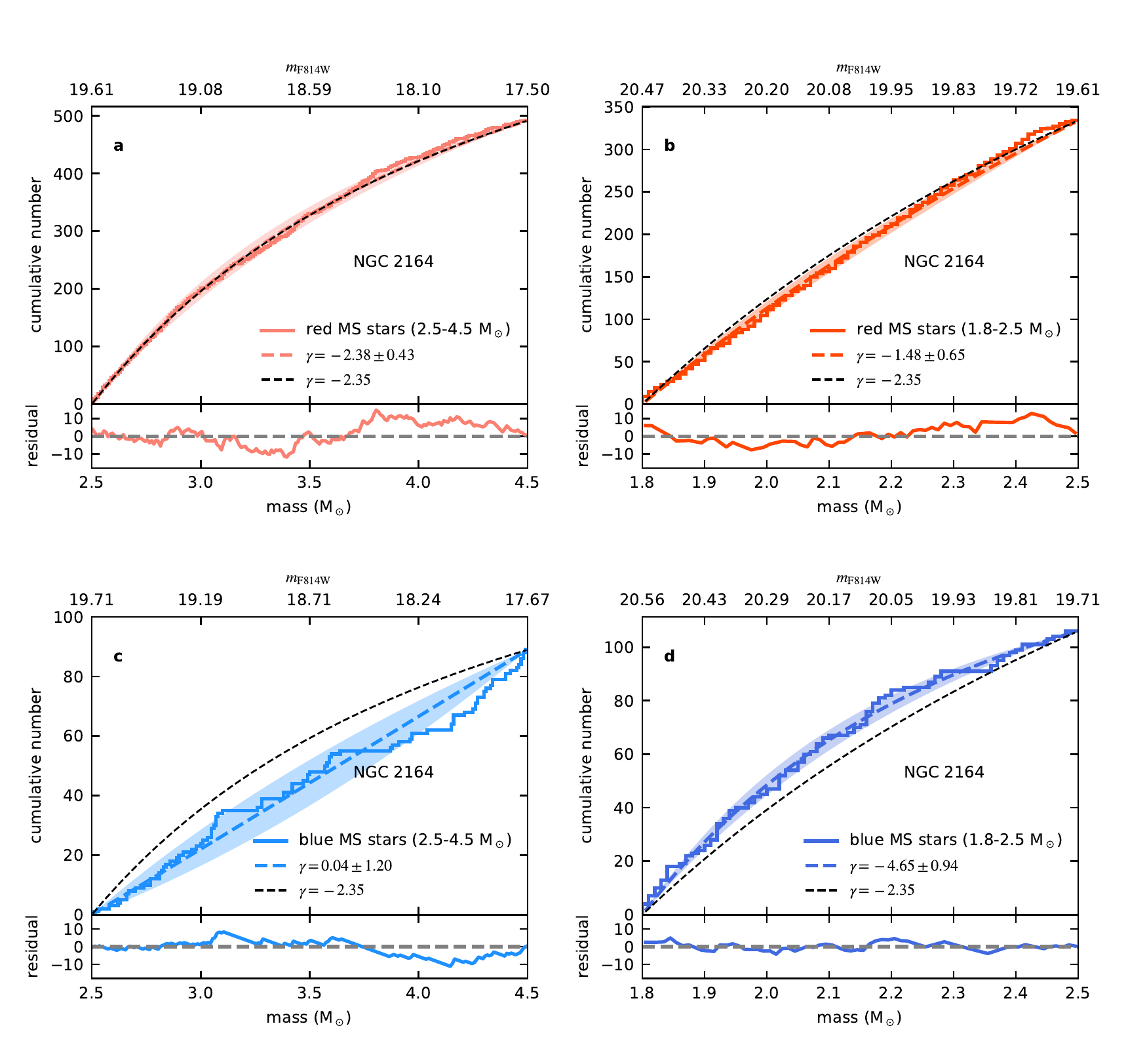}
\caption{Mass function of the red and the blue main-sequence stars in NGC\,2164. The plots are the same as Supplementary Figure\,11.
}
\label{Suppfig:n2164_MF}
\end{figure}     

\begin{figure}[ht]
\centering
\includegraphics[width=\linewidth]{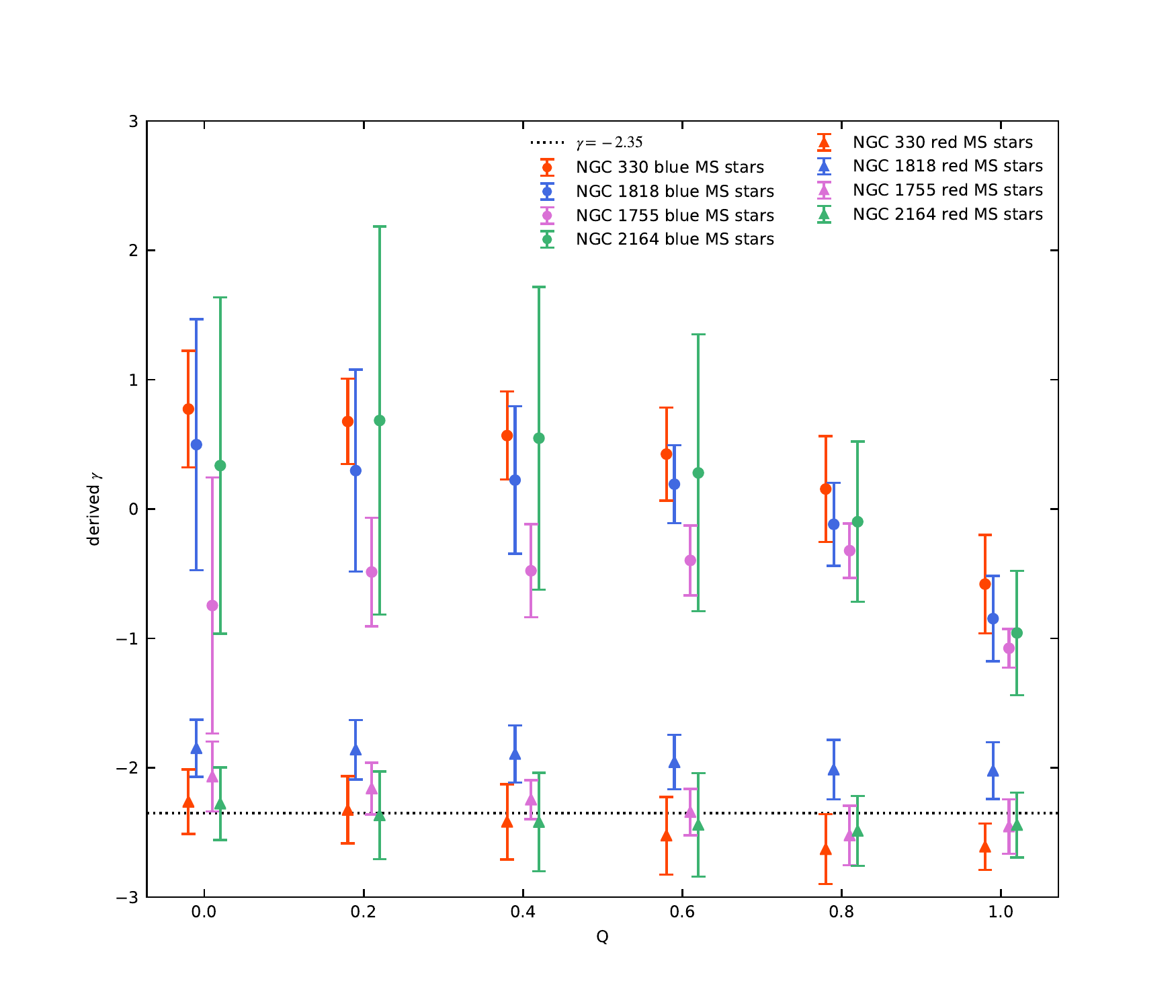}
\caption{Dependence of the mass function slope on the borderline between blue and red main-sequence stars. The borderline is obtained by shifting the isochrone for the blue main sequence by $Q$ times the color separation between the isochrones for the blue and the red main sequences. We consider $Q$ values between 0 and 1, in intervals of 0.2. The filled circles and triangles with error bars denote the derived mass function slope for the blue and the red main-sequence stars with their 1$\sigma$ errors, respectively. Different colors correspond to different clusters. We avoid to plot the errorbars at the same $Q$ values to make them distinct. The black dashed line marks the slope of the Salpeter IMF $\gamma=-2.35$.
}
\label{Suppfig:MF_diff_Q}
\end{figure}

\section*{E: Merger models and merger history}\label{suppsec:E}
We suggest that MS mergers are responsible for the formation of the blue MS stars, according to the fact that they can produce blue stragglers \cite{2020ApJ...888L..12W} that appear younger than the other cluster stars. In this scenario, we can estimate the possible merger time $t_\mathrm{merge}$ of each blue MS star, which is the time in the cluster history when the merger happens,  by comparing its distribution in the CMD with the theoretical merger models. We next describe in detail how we do this.

For each initial rotation, we construct a series of lines in the CMD that represent the current positions of the merger products whose progenitors coalesced at given times in the past, from 0\,Myr after starburst to the age of the cluster, in intervals of 1\,Myr. Supplementary Figure\,14 shows some representative lines by assuming rotational velocities of 35\% of the break-up velocities for the merger products. 
We then use the nearest line to determine $t_\mathrm{merge}$ for each observed blue MS star. 
The adoption of faster rotating merger models results in a larger $t_\mathrm{merge}$ for a given observed blue MS star, because faster rotating merger models are redder than the slower rotating ones with the same $t_\mathrm{merge}$. 
Supplementary Figure\,14 indicates that most of the observed blue MS stars above 19.5th magnitude can be covered by merger models with a rotation parameter of $W_{\rm i} = 0.35$, however, there are outliers
to the right and left of the family of lines, indicating that somewhat larger or slower rotation is required for these cases.
For example, the blue MS stars between the two isochrones that are used to fit the observed blue and red MS stars (see Fig.\,1b) cannot be reproduced by merger models with a rotational rate smaller than 35\% of their critical velocities. For those stars, we calculate the maximum and minimum rotational rates with which merger models can reach these stars in the CMD.

The derived $t_\mathrm{merge}$ for the NGC\,1755 blue MS stars are shown in Fig.\,4b (lower panel), with the lower limits determined by the merger models with the maximum allowed rotation and the upper limits determined by the merger models with the minimum allowed rotation. 
We only include stars brighter than 19th magnitude, because we lose diagnostic power beyond that as the fastest rotating merger models produced at 58\,Myr are bluer than the slowly rotating merger models produced at 0\,Myr. 
We also exclude the brightest blue MS stars whose magnitudes are smaller than 15.5, because they are brighter than all our merger models. Perhaps they are extremely fast rotating stars which may be subject to strong gravity darkening/brightening.  
Finally, we compute the merger history by assuming a Gaussian distribution within the allowed merger velocities, with a mean value of $W_{\rm i} = 0.35$ and a standard deviation of $W_{\rm i} = 0.2$. 
This distribution is chosen because the isochrone with $W_{\rm i}=0.35$ can fit the blue MS well, and the width of this Gaussian distribution can cover all the identified blue MS stars. This leads to the merger event history of the blue MS stars of NGC\,1755 as shown in Fig.\,4a (upper panel). The time resolution is 1\,Myr, which is governed by the $t_\mathrm{merge}$ interval used when we build the merger models. 
We perform a bootstrapping analysis to obtain the uncertainty of the derived merger time distribution. We randomly assign a rotational velocity for each blue MS star following a Gaussian distribution, and derive its corresponding merger time. We repeat this process 10\,000 times, and determine the $1\sigma$ uncertainty as the place where 68\% of the 10\,000 obtained number of mergers per Myr is included in each $t_\mathrm{merge}$ bin. The result is shown by the shaded area in Fig.\,4a.
Our results reveal that merger events should be prevalent in the first tens of Myrs, with a peak at 0\,Myr to 2\,Myrs, to account for the many observed stars near the blue MS.

The distribution of the blue MS stars in the CMD indicates an earlier merger time for brighter stars (see the blue MS stars with magnitudes between 16.5 and 18 near the $t_\mathrm{merge}=0$ line in Supplementary Figure\,14). In order to examine the correlation between the derived merger time and magnitude, we calculate the Spearman's rank correlation coefficient between these two variables in all 10\,000 bootstrapping simulation. The obtained coefficient ranges from 0.38 to 0.60, with an average value of 0.50, indeed indicating a moderate positive correlation.  Such a correlation may be consistent with the results of recent binary orbit decay simulations, which propose that more massive binaries are expected to merge earlier than binaries with lower massive stars due to their larger angular momentum and energy loss rates caused by dynamical friction \cite{2012A&A...543A.126K}.

In order to check whether the derived merger rate history is affected by our isochrone fitting, we do the same experiment using the isochrones shown in Supplementary Figure\,7. The results are shown in Supplementary Figure\,15. This time, we consider the merger products to have 0 to 45\% of their critical velocities, following a Gaussian distribution with a peak at $W_{\rm i} = 0.15$ and a width of $W_{\rm i}=0.3$. These values are chosen such that the rotation peak matches the rotation of the single star models that are used to fit the blue MS, and all blue MS stars are covered, except for several stars which fall on the blue side of the zero-age MS line in Panel\,a of Supplementary Figure\,7. The rise of the merger event rate at recent times in Supplementary Figure\,15 is caused by ample blue MS stars that can only be reached by non/slowly rotating mergers formed very recently. These blue MS stars have smaller derived rotational rates and $t_\mathrm{merge}$ range, thus a larger probability in each allowed $t_\mathrm{merge}$ bin. However, our main conclusion that merger events are prevalent in the first tens of Myr remains intact. The Spearman's rank correlation coefficient in this case is 0.59.

We use the same method to estimate the merger history of the blue MS stars in the SMC cluster NGC\,330 and the LMC clusters NGC\,1818 and NGC\,2164, with the results shown in Supplementary Figure\,16. The isochrone fitting is done by employing $W_{\rm i}=0.65$ and $W_{\rm i}=0.35$ models. 
We see that our conclusion that mergers happen frequently in the first tens of Myr holds in all the studied clusters. Besides, the derived merger event frequencies in these three clusters all show a continuously decreasing trend. The mean derived Spearman's rank correlation coefficients are 0.20, 0.59 and 0.48 for clusters NGC\,330, NGC\,1818 and NGC\,2164, respectively, all implying a moderate positive correlation between the derived merger time and magnitude.

In the above analysis, we only consider binary mergers from equal-mass binaries, which possess the strongest rejuvenation. In the following, we examine how the mass ratio of the merger progenitors affects our results.
We first quantify the relation between the apparent rejuvenation (how much younger the star looks than it really is) and the mass ratio. We describe the apparent rejuvenation as ($t-t_{\rm app}$)/$t$, where $t$ is the time at which a merger happens, and $t_{\rm app}$ is the apparent age of the merger product. We consider binaries with primary masses between $2.1\,\mso$ and $10\,\mso$ in intervals of $0.5\,\mso$, and mass ratios between $0.2$ and $1$ in intervals of $0.2$.
We assume mergers happen at 58\,Myr (i.e., the derived age of cluster NGC\,1755), and show the result in Supplementary Figure\,17.
It can be seen that rejuvenation increases with increasing mass ratio. For example a 5\,$\mso$ merger product can be
rejuvenated to $\sim 40\%$ and $\sim 80\%$ younger than its progenitors if it is produced by a $q=0.2$ and a $q=1$ binary, respectively.
In order to perform this examination, we extend our single star models to 0.4\,$\mso$. But we only build and use the zero-age MS models for such low-mass stars, as they hardly evolve in young star clusters.

We show the distribution of our merger products created at 58\,Myr from binaries with different mass ratios in the CMD in Supplementary Figure\,18, and the resulting merger history in Supplementary Figure\,19. We only include blue MS stars that can be covered by our merger models in this check. 
It leads to different integral number of the derived merger events under the assumption of different mass ratios. The merger models constructed with low mass ratios are redder in the CMD compared to their high mass-ratio counterparts, thus can cover fewer observed blue MS stars.  
In Supplementary Figure\,18 and Supplementary Figure\,19, we see that results for mass ratios larger than 0.6 are nearly identical. 
The rate of the merger events increases after $\sim$ 54\,Myr when adopting mass ratios smaller than 0.4, because more blue MS stars have smaller derived rotational rate and merger time ranges.
Nevertheless, our main conclusion that merger events should occur at early times is robust, regardless of the binary mass ratio.

\begin{figure}[ht]
\centering
\includegraphics[width=0.8\linewidth]{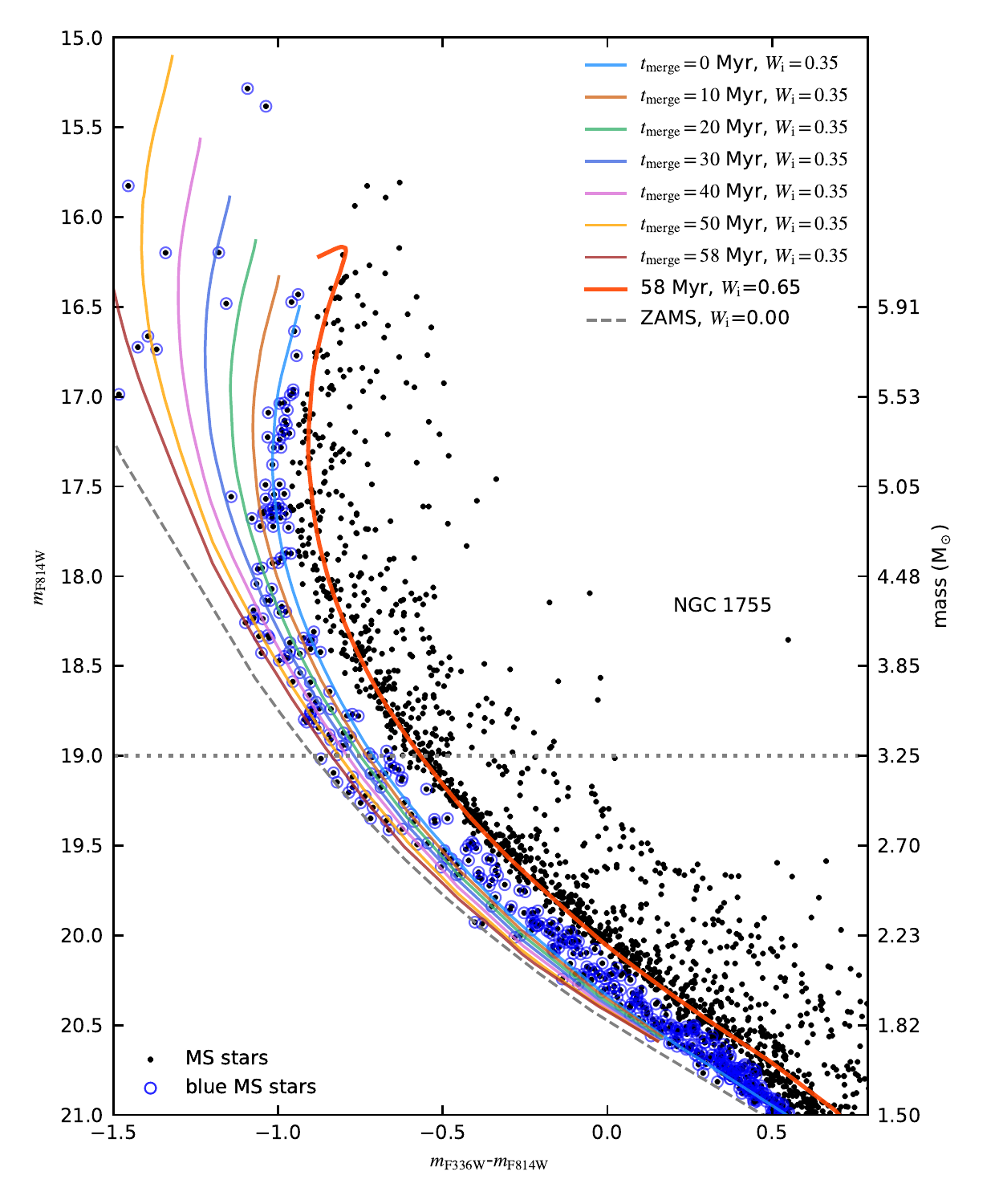}
\caption{Lines of constant merger time in the color-magnitude diagram. The distribution of the main-sequence stars in NGC\,1755 is shown, with blue open circles indicating blue main-sequence stars.
Solid lines of given colors provide the location of merger products which formed
from equal-mass binaries at the indicated time (see color scale) 
with 35\% of critical rotation,
and were then evolved to the current cluster age. 
The blue main-sequence stars above the grey horizontal dotted line are considered in the merger time estimation (see Supplementary Information E).
The isochrone fit to the red MS (solid red line), and the resulting zero-age main-sequence line (dashed grey), are as in Fig.\,1b.  
}
\label{fig:n1755_merger_iso}
\end{figure}

\begin{figure}[ht]
\centering
\includegraphics[width=\linewidth]{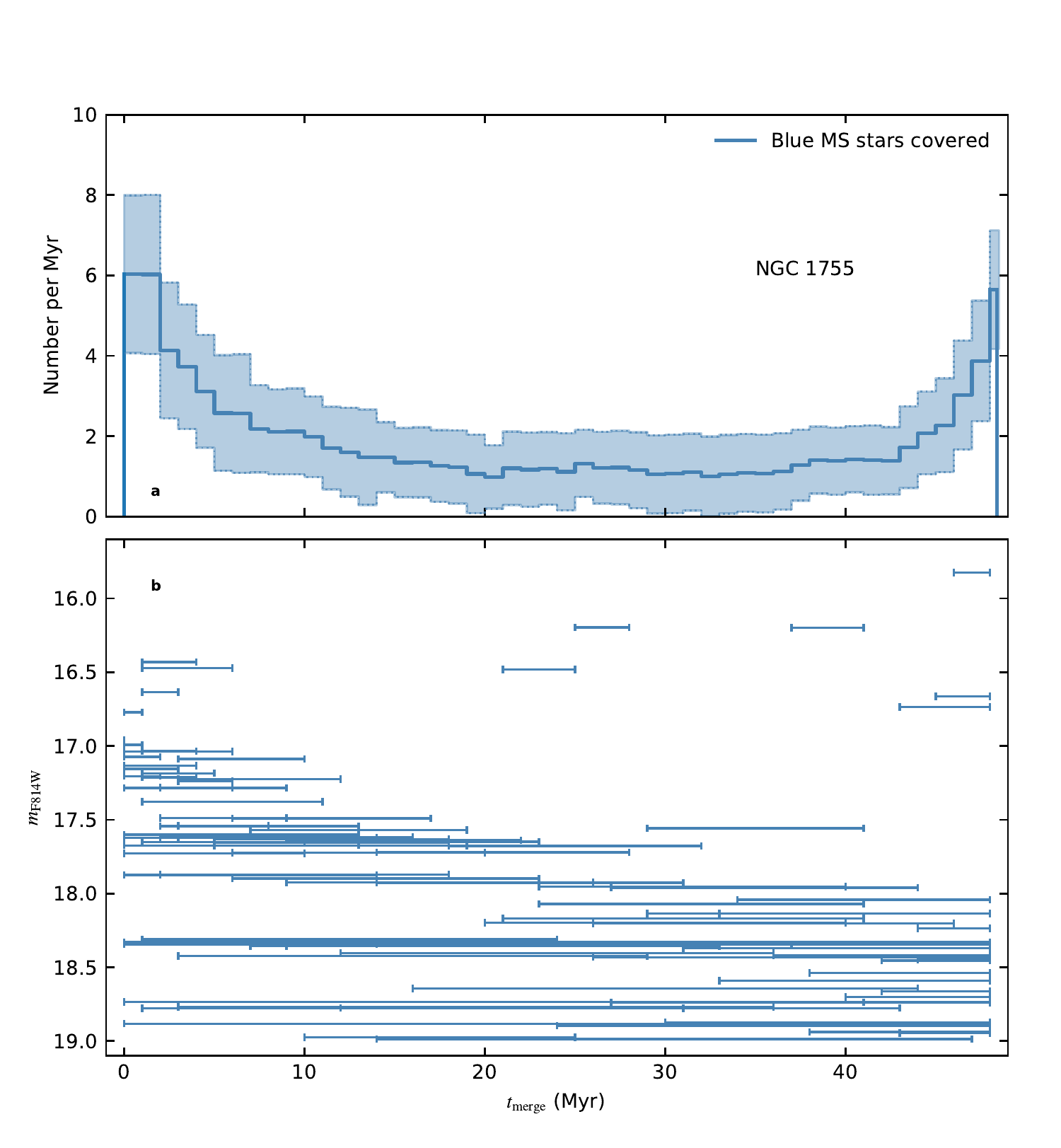}
\caption{Merger history of the blue main-sequence stars in NGC\,1755 based on alternative initial rotational velocities (see Supplementary Figure\,7a). The plots are the same as Fig.\,4.
}
\label{Suppfig:n1755_merger_time_v55}
\end{figure}

\begin{figure}[ht]
\centering
\includegraphics[width=\linewidth]{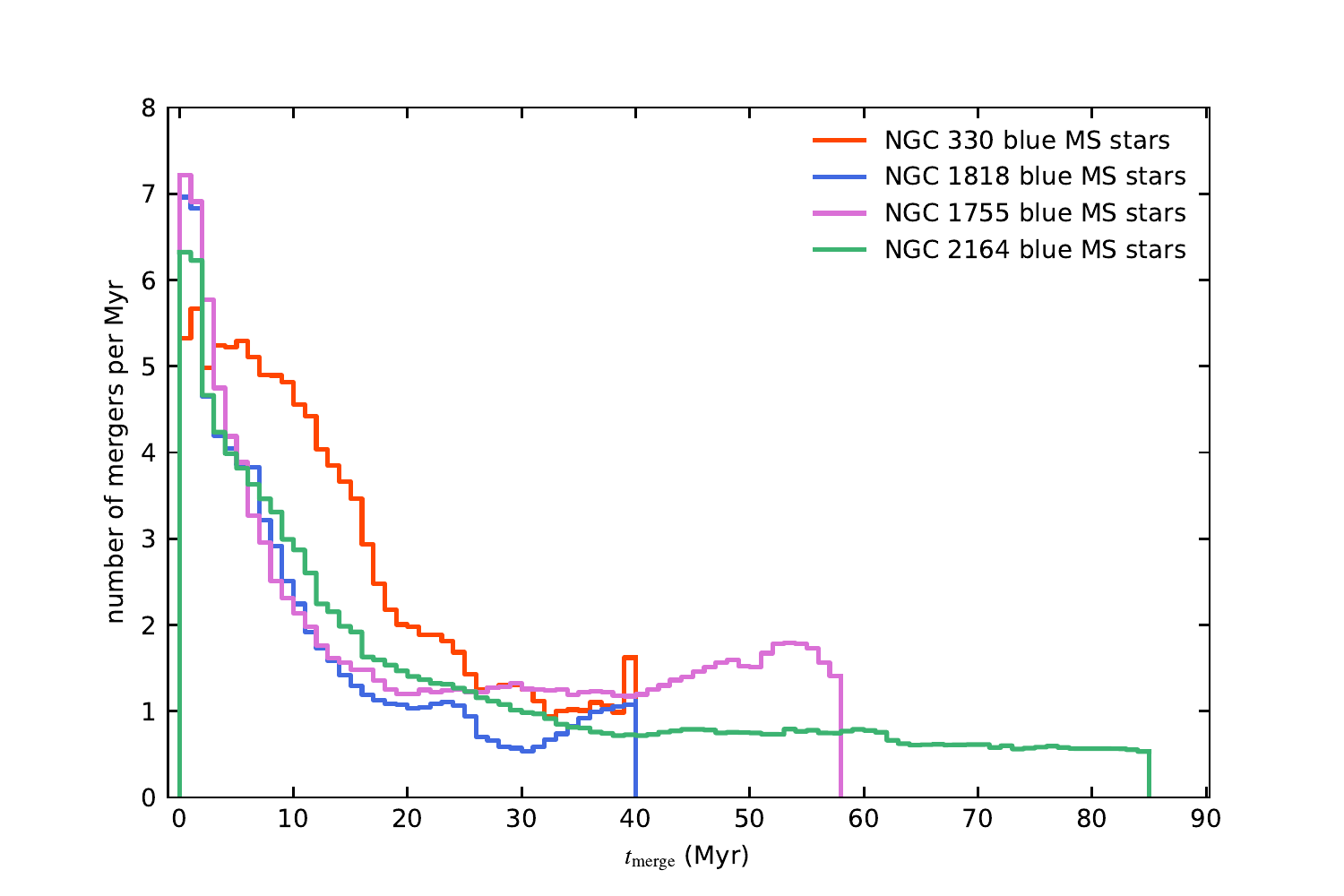}
\caption{Merger history of the blue main-sequence stars in four young Magellanic Cloud clusters. Different colors correspond to different star clusters.}
\label{Suppfig:tmerg_diff_cluster}
\end{figure}    

\begin{figure}[ht]
\centering
\includegraphics[width=\linewidth]{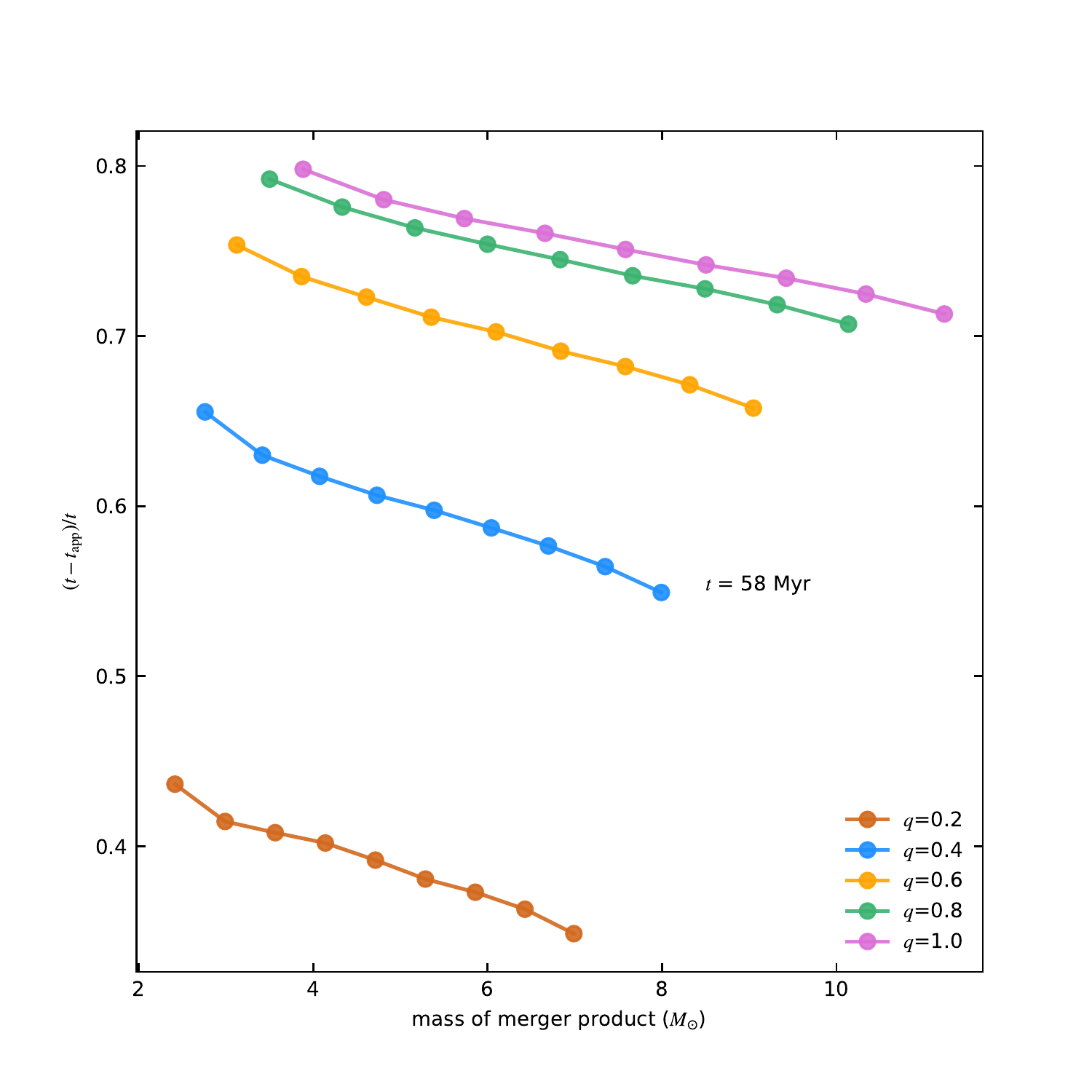}
\caption{Relative rejuvenation for different post-merger stellar masses and pre-merger mass ratios. $t$ is the cluster age of NGC\,1755, which also indicates the most recent incidence time of the merger. $t_{\rm app}$ is the apparent age of the merger product as measured by
single star isochrones, for a fixed merger time of $t=58$\,Myr. The five lines correspond to five different mass ratios
as indicated by their color (see legend). Each dot on the line corresponds to one computed binary model (see Supplementary Information\,E).
}
\label{Suppfig:rejuv}
\end{figure}     

\begin{figure}[ht]
\centering
\includegraphics[width=0.8\linewidth]{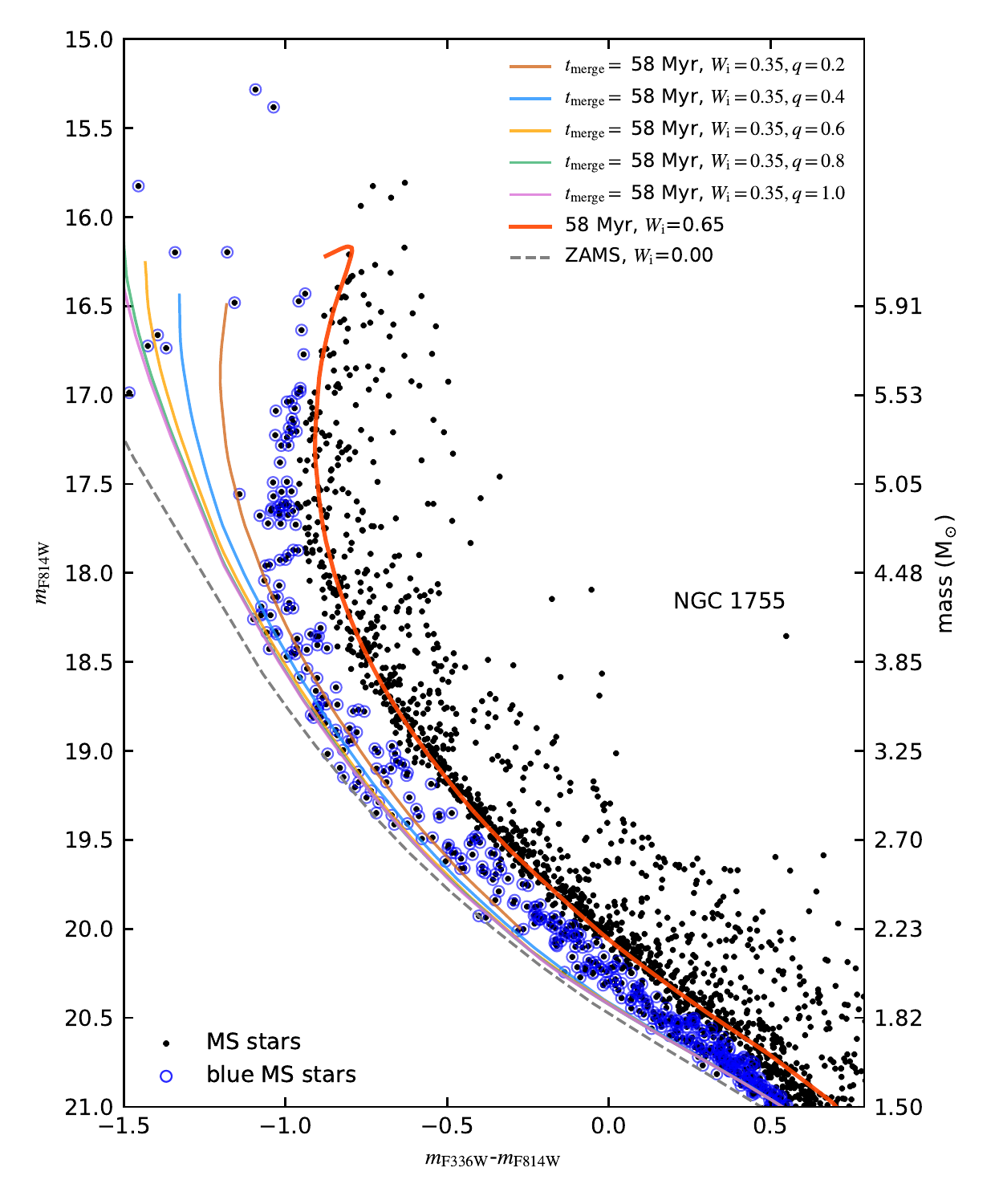}
\caption{Impact of the pre-merger mass ratio on the position of the merger product in the color-magnitude diagram. The thin solid lines correspond to the merger models rotating at 35\% of their critical velocities. Different colors correspond to different mass ratios for the precursor binary models. We only show models of mergers for a merger incidence time of 58\,Myr. The isochrone for the red main sequence (thick solid red line), and the zero-age main-sequence line (dashed grey) are the same as in Fig.\,1b.
}
\label{Suppfig:n1755_merg_iso_diff_q}
\end{figure}        

\begin{figure}[ht]
\centering
\includegraphics[width=\linewidth]{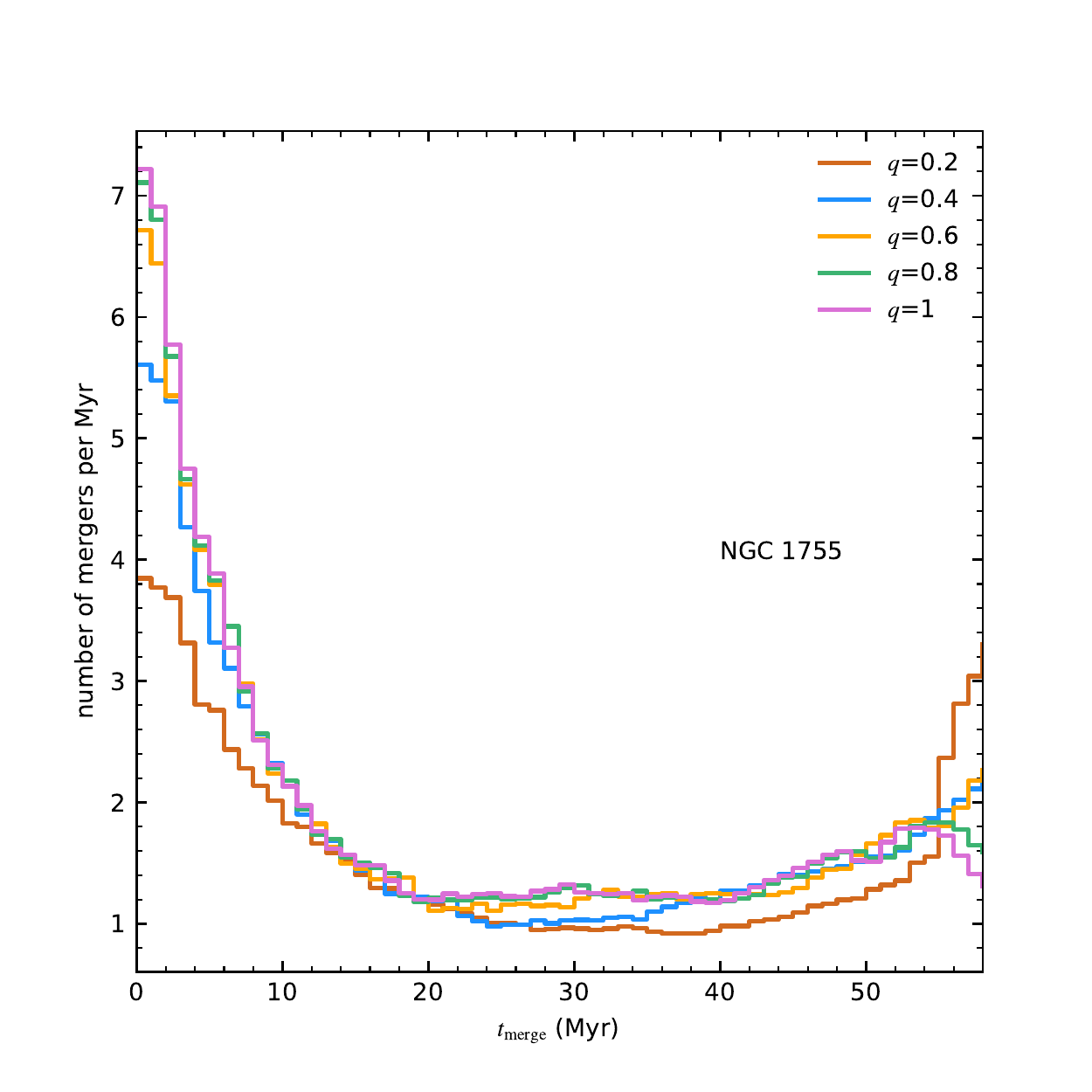}
\caption{Effect of the pre-merger mass ratio on the derived merger history. Different colors show the results of the binary merger models with different initial mass ratios. 
}
\label{Suppfig:tmerg_diff_q}
\end{figure}         
       
Apart from the above-mentioned uncertainties, the derived merger times may also be affected by additional mixing during the merger process, which can make a merger product appear even younger (1.14 times for stars above 5$\mso$ and 1.43 times for stars below 5$\mso$) than in the case of fully rejuvenation \cite{2008A&A...488.1017G,2013MNRAS.434.3497G}. However, this additional mixing does not significantly affect the distribution of early mergers in the CMD, as the long-lasting later evolution of the merger product after it has been created washes out this small natal difference. Whereas potential surface He enhancement may also impact the distribution of the merger products, it is not found in both theoretical simulations or spectroscopic observations \cite{2001ApJ...548..323S,2013MNRAS.434.3497G,2020AJ....159..152C}.

We have proposed in the main text that orbital decay in multiple systems happens probably via Kozai-cycles. A glimpse of the MS distribution in the CMD at the present clusters gives us the impression that there are many suspected triples or higher-order multiples (see Fig.\,1b and Supplementary Figure\,5, those with colors redder than the equal-mass binary line red shifted by three times the photometric error). To indicate this, we derive the fraction of the suspected triple or higher-order multiples with respect to all MS stars in three clusters, one SMC cluster NGC\,330, and two LMC clusters, NGC\,1818 and NGC\,1755. 
In figures comparing our theoretical isochrones with the observations in the CMD (Fig.\,1b and Supplementary Figure\,5), we count the stars on top of the equal-mass binary lines shifted redwards by three times the photometric errors as the suspected triple or higher-order multiples. 
The results are shown in Supplementary Figure\,20.  We use absolute magnitude such that we can directly compare clusters with different ages in different galaxies. We use isochrones of $W_{\rm i}=0.65$ single star models that can best fit the red MS in each cluster to convert the apparent magnitude to the absolute magnitude. We consider stars whose absolute magnitudes are 2 mags larger than the turn-off magnitude to exclude the probable contaminations from critically rotating Be stars. At the low-brightness end, we cut at $M_{\rm F814W}=2$, which roughly corresponds to an apparent magnitude of 20.5 in these clusters.
We show our results in different magnitude intervals. The error bars on the x-axis reflect the magnitude intervals, while the error bars on the y-axis reflect the Poisson error.
In general, the fraction of the suspected triples decreases with increasing magnitude.  We point out that these values should only be the lower limit, as we miss the low mass ratio multiple systems whose positions are below the boundary lines.

 \begin{figure}[ht]
\centering       
\includegraphics[width=\linewidth]{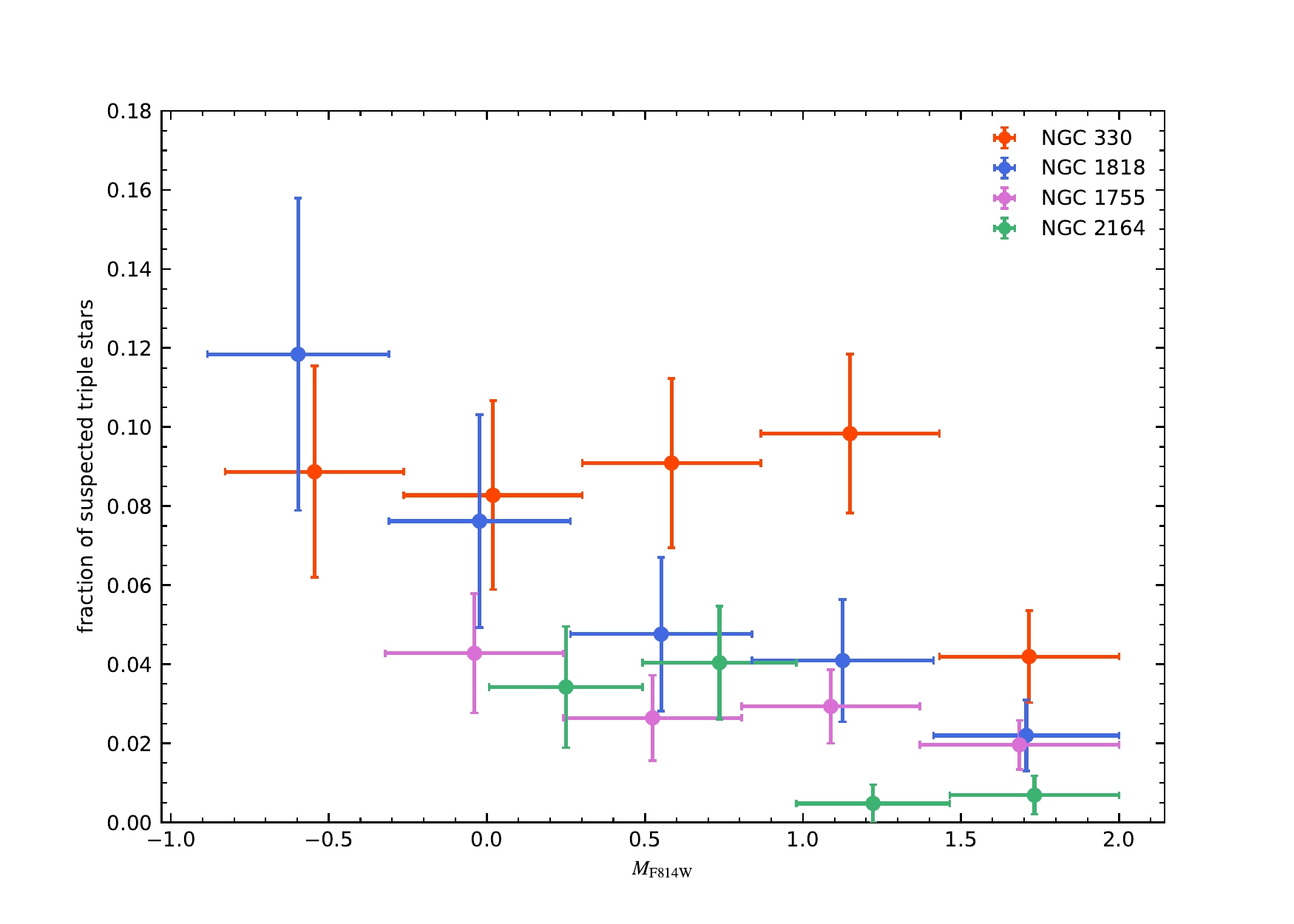}
\caption{Fraction of triple stars (or higher order multiples) in four young Magellanic Cloud clusters. The estimations are based on the equal-mass binary lines in the color-magnitude diagrams (see Fig.1b, and Supplementary Figure\,5). We consider absolute magnitude for stars in four young Magellanic Cloud clusters in different magnitude intervals. The horizontal width of the error bars correspond to the magnitude intervals, while the vertical error bars reflect the Poisson error.}
\label{Suppfig:triple}
\end{figure}  
   
\bibliography{scibib}
\bibliographystyle{naturemag}

\end{document}